\documentclass[aps,pre,twocolumn,showpacs,preprintnumbers,superscriptaddress,amsmath,amssymb]{revtex4-1}

\usepackage[T1]{fontenc}
\usepackage[latin1]{inputenc}
\usepackage{graphicx}
\usepackage{amsfonts, amsmath}
\usepackage{amstext}
\usepackage{color}
\usepackage{dsfont}
\usepackage{subfig}

\newcounter{commentzaehler}
\usepackage{adjustbox}
\begin{document}
%\graphicspath{{figs/}}
%\preprint{LEH14, dated \today}
\hyphenation{syn-chro-ni-za-tion}
\title{Controlling cluster synchronization by adapting the topology}

\author{Judith Lehnert}
\email[corresponding author:]{lehnert@itp.tu-berlin.de}
\affiliation{Institut f{\"u}r Theoretische Physik, Technische 
Universit{\"a}t Berlin, Hardenbergstra{\ss}e 36, 10623 Berlin, 
Germany}
\author{Philipp H\"ovel}
\affiliation{Institut f{\"u}r Theoretische Physik, Technische 
Universit{\"a}t Berlin, Hardenbergstra{\ss}e 36, 10623 Berlin, 
Germany}
\affiliation{Bernstein Center for Computational Neuroscience, 
Humboldt-Universit{\"a}t zu Berlin, Philippstra{\ss}e 13, 10115 
Berlin, Germany}
\author{Anton~Selivanov}
\affiliation{Department of Theoretical Cybernetics, Saint-Petersburg State University, Saint-Petersburg, Russia}
\author{Alexander~Fradkov}
\affiliation{Department of Theoretical Cybernetics, Saint-Petersburg
  State University, Saint-Petersburg, Russia}
\affiliation{ Institute for Problems
of Mechanical Engineering, Russian Academy of Sciences,
Bolshoy~Ave,~61, V.~O.,~St.~Petersburg, 199178~Russia }

\author{Eckehard Sch\"{o}ll}
\affiliation{Institut f{\"u}r Theoretische Physik, Technische
  Universit{\"a}t Berlin, Hardenbergstra{\ss}e 36, 10623 Berlin, Germany}

% Synchronization, nonlinear dynamics, 05.45.Xt
% neural networks, 87.85.dq
% Delay equations, in function theory, 02.30.Ks
% Complex systems, 89.75.-k

\begin{abstract} We suggest an adaptive control scheme for the control
of in-phase and cluster synchronization in delay-coupled
networks. Based on the {\it speed-gradient method}, our scheme adapts
the topology of a network such that the target state is realized. It is
robust towards different initial condition as well as changes in the
coupling parameters. The emerging topology is characterized by a
delicate interplay of excitatory and inhibitory links leading to the
stabilization of the desired cluster state. As a crucial parameter
determining this interplay we identify the delay time. Furthermore, we
show how to construct networks such that they exhibit not only a given
cluster state but also with a given  oscillation frequency.
We apply our method to coupled Stuart-Landau
oscillators, a paradigmatic normal form that naturally arises in an
expansion of systems close to a Hopf bifurcation. The successful and
robust control of this generic model opens up possible applications in
a wide range of systems in physics, chemistry, technology, and life
science.
\end{abstract}

\maketitle

\section{Introduction } Networks play a prominent role in the research of
very different fields, ranging from social science, economics, and 
psychology to biology, physics, and mathematics
\cite{BOC06a,ALB02a,NEW03}, as they are a straight-forward concept to
describe the interactions of many systems or agents. While previous
research focused either on the formation of topologies
\cite{WAT98,BOC06a,ALB02a,NEW03,RAP57,ERD59} or on the dynamics on a
network with fixed topology
\cite{DHA04,ZIG09,CHO09,CHA05,SOR07,ALB00,KIN09,LEH11,KEA12}, adaptive
networks bring these two aspects together: In such networks the
topology evolves according to the state of the system while the
dynamics on the network and thus the state is influenced by the
topology \cite{GRO08a}.

%\comment{maybe a schematic diagram how network topology and dynamics 
%interact?}

Among the variety of well-studied dynamical scenarios, in-phase 
synchronization has been  the main focus of
research concerning the dynamics on networks for a long time. Recently, more 
complex 
synchronization patterns, including cluster and group synchronization 
have received growing interest both in theory 
\cite{SOR07,DAH12,KAN11,KAN11a,GOL02a, SOR14} and in  experiments 
\cite{BLA13,WIL13}.
These scenarios appear in many biological systems, examples include
dynamics of nephrons \cite{MOS02}, central pattern
generation in animal locomotion \cite{IJS08}, or population dynamics
\cite{BLA99a}. %Potential application are in communication 
In an $M$-cluster state, for instance, the compound system evolves to 
$M$ clusters with in-phase synchronization between the nodes of one 
cluster and  -- in the case of an oscillatory system -- with a constant 
phase lag of $2\pi/M$ between the clusters. In contrast, group synchronization 
refers to the case where each cluster potentially exhibits different 
local dynamics.  

Control of cluster synchronization by adaptively changing
the topology of the network has previously been investigated, to our 
knowledge, only by a few researchers: Lu et al. consider control of 
cluster synchronization by means
of changing topology. As a limiting restriction for the 
applicability, their method requires a-priori knowledge to which 
cluster each node should belong in the final state
\cite{LU09}. Furthermore, the majority of algorithms developed to control (mainly
in-phase) synchronization by adaptation of the network topology are 
based on local mechanisms. Most of them are related to Hebb's rule: 
{\it Cells that fire together, wire together} \cite{HEB49}. Our method 
uses a global goal function to realize self-organized
control and is therefore a powerful alternative and complements 
existing control schemes.

In Ref.~\cite{SEL12}, the control of cluster synchronization adapting 
the phase
of a complex coupling strength was shown. While this method is an 
easy and
elegant way to control small networks by tuning of just one parameter, 
it fails for
larger networks because it becomes too sensitive to initial conditions.  

Here, we present an algorithm to adapt the topology of a network in
such a way that a desired cluster state is reached. The method is not
only robust towards initial conditions but also works for a large
parameter range. The adaption algorithm for the network 
structure employs the \textit{speed-gradient method} 
\cite{FRA07}. As a strong advantage compared to other methods, no a-priori ordering of nodes is needed, i.e., it is
not necessary to assign each node to a specific cluster in advance. 
In contrast, the final assignment is a built-in consequence of the 
initially designed goal function.

Aiming for a general framework, we consider the Stuart-Landau 
oscillator, which is the normal form
of a Hopf bifurcation and therefore generic for many oscillating
systems present in nature and technological applications. In 
addition, we implement time-delayed coupling between the nodes 
because delays naturally arise in many applications. Note that our 
scheme also works for instantaneous coupling.

Furthermore, it is not necessary to control all links of a network. 
The control scheme remains successful if only a 
subset of links
is accessible. We will explicitly demonstrate this by 
applying the method to parts of a random network \cite{ERD60}, while retaining 
control of the target state.

To what extent structure determines function, i.e., which topologies
allow for cluster synchronization, is a hot topic of current research
\cite{KAN11, KAN11a, GOL02a}. Our method provides an easy and
self-organized way to generate weighted networks that are able to
exhibit cluster synchronization. The topology of these networks will
contain some randomness as we start with random initial conditions for
the state of the node. However, on average the topology is
characterized by common features that enable synchronization and hence
yield the desired cluster state. As a crucial parameter shaping this
topology we identify the delay time. Furthermore, we show how to
construct networks in which cluster states oscillate with a desired
frequency (including zero frequency).

In the following Section, we introduce the model and establish the
target states. Section~\ref{sec:sgm} discusses the goal function and briefly reviews the
speed-gradient method. The central part of the paper is
Sec.~\ref{sec:adapting-topology-}, in which the  adaption algorithm for the
topology is developed and demonstrated. The structure of the networks after
successful control is discussed in
Sec.~\ref{sec:struct-prop-}, where we also show how a discrete Fourier
transform of the coupling matrix can be used to get insight into
the delay-modulated topology.  Section~\ref{sec:choosing-frequency-}
shows how the obtained results can be used to not only control
cluster synchronization, but also to grow networks in which a cluster state
with a given frequency is stable. We conclude with  Sec.~\ref{sec:conclusion-}.   

%%%%%%%%%%%%%%%%%%%%%%%%%%%%%%%%%%%%%%%%%%%%%%%%%%%%%%%%%%%%%%%%%%%%%%%%%%%%%%%%%%%%%%%%%%%%%%%%%
\section{Model} \label{sec:model}
The system equation of a Stuart-Landau oscillator is given by
\begin{equation}
    \label{einsl}
    \dot{z}= \left[\lambda + i\omega - |z|^2\right]z,
\end{equation}
with the complex variable $z \in \mathds{C}$ and $\lambda,\omega \in
\mathds{R}$ \cite{KUR84}. It arises in a center
manifold expansion
close to a supercritical Hopf bifurcation with $\lambda$ as the  bifurcation
parameter.
% With the $-$ and $+$ sign, Eq.~\eqref{einsl} describes supercritical or subcritical
%Hopf bifurcation, respectively. In the following, only the
%supercritical case will be considered:
For $\lambda<0$, the system exhibits a stable focus. At the
bifurcation point $\lambda=0$, the
focus loses stability, and a stable limit cycle with radius
$\sqrt{\lambda}$ is born. The parameters
$\omega$  denotes the frequency of the uncoupled node.

In the following, we consider a system of $N$ delay-coupled
Stuart-Landau oscillators $z_j$, $j=1,\dots,N$:
\begin{eqnarray}\label{sl}
  \dot{z}_j(t) &=& [\lambda + i\omega - |z_j|^2]z_j\nonumber\\
&&+K\sum_{n=1}^N G_{jn}(t)[z_n(t-\tau)-z_j(t)],
\end{eqnarray}
where $K$ is the overall coupling strength, and $\tau$ is the delay time. In the
following, we denote
delayed variables by an index $\tau$, e.g.,  $z_\tau \equiv
z(t-\tau)$. $\{G_{jn}(t)\}_{j,n=1,\ldots,N}$ is the coupling matrix
describing the
topology of the network and subject
to the adaptive control.

Equation~\eqref{sl} can be rewritten in amplitude and
phase variables with  $r_j=|z_j|$ and $\varphi_j=\arg(z_j)$:

\begin{eqnarray} 
  \dot{{r_j}}(t)&=&\left[\lambda - r_j^2\right]r_j+K\sum_{n=1}^N{G_{jn}\left\{ r_{n,\tau}\cos\left[\varphi_{n,\tau}-\varphi_j\right]-r_{j}\right\}},\nonumber\\
  \dot{{\varphi_j}}(t) &=&\omega+K\sum_{n=1}^NG_{jn}\left\{\frac{r_{n,\tau}}{r_j}\sin\left[\varphi_{n,\tau}-\varphi_j\right]\right\}.
  \label{eq:sl_r}
\end{eqnarray}
%where we denoted the time-delayed variables by an index $_\tau$.

\begin{figure}
\begin{picture}(0,0)%
\includegraphics{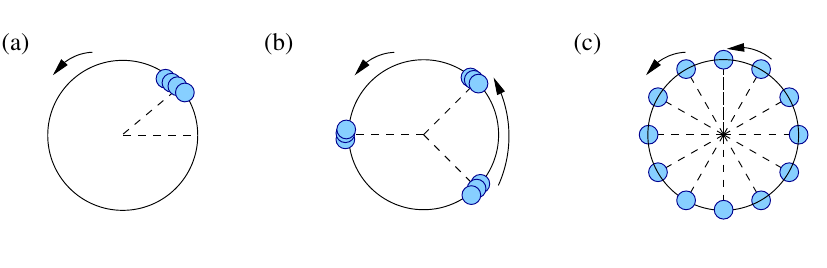}%
\end{picture}%
\setlength{\unitlength}{2368sp}%
\begingroup\makeatletter\ifx\SetFigFont\undefined%
\gdef\SetFigFont#1#2#3#4#5{%
  \reset@font\fontsize{#1}{#2pt}%
  \fontfamily{#3}\fontseries{#4}\fontshape{#5}%
  \selectfont}%
\fi\endgroup%
\begin{picture}(6585,2129)(-389,-16123)
%  0 
\put(3731,-15151){\makebox(0,0)[lb]{\smash{{\SetFigFont{9}{10.8}{\rmdefault}{\mddefault}{\updefault}{\color[rgb]{0,0,0}$\frac{2 \pi} {3}$}%
}}}}
%  0 
\put(1024,-14576){\makebox(0,0)[lb]{\smash{{\SetFigFont{9}{10.8}{\rmdefault}{\mddefault}{\updefault}{\color[rgb]{0,0,0}$z_j$}%
}}}}
%  0 
\put(5491,-14201){\makebox(0,0)[lb]{\smash{{\SetFigFont{9}{10.8}{\rmdefault}{\mddefault}{\updefault}{\color[rgb]{0,0,0}$\frac{2 \pi} {N}$}%
}}}}
%  0 
\put(6006,-15571){\makebox(0,0)[lb]{\smash{{\SetFigFont{9}{10.8}{\rmdefault}{\mddefault}{\updefault}{\color[rgb]{0,0,0}$z_N$}%
}}}}
%  0 
\put(6181,-15151){\makebox(0,0)[lb]{\smash{{\SetFigFont{9}{10.8}{\rmdefault}{\mddefault}{\updefault}{\color[rgb]{0,0,0}$z_1$}%
}}}}
%  0 
\put(6081,-14761){\makebox(0,0)[lb]{\smash{{\SetFigFont{9}{10.8}{\rmdefault}{\mddefault}{\updefault}{\color[rgb]{0,0,0}$z_2$}%
}}}}
%  0 
\put(4606,-15591){\makebox(0,0)[lb]{\smash{{\SetFigFont{9}{10.8}{\rmdefault}{\mddefault}{\updefault}{\color[rgb]{0,0,0}$z_m$}%
}}}}
%  0 
\put(1726,-15136){\makebox(0,0)[lb]{\smash{{\SetFigFont{9}{10.8}{\rmdefault}{\mddefault}{\updefault}{\color[rgb]{0,0,0}$z_{2+3i}$}%
}}}}
%  0 
\put(3478,-15741){\makebox(0,0)[lb]{\smash{{\SetFigFont{9}{10.8}{\rmdefault}{\mddefault}{\updefault}{\color[rgb]{0,0,0}$z_{3+3i}$}%
}}}}
\put(1726,-16036){\makebox(0,0)[lb]{\smash{{\SetFigFont{9}{10.8}{\rmdefault}{\mddefault}{\updefault}{\color[rgb]{0,0,0}$(i=0, \dots, (N-1)/3 )$}%
}}}}
\put(-224,-16036){\makebox(0,0)[lb]{\smash{{\SetFigFont{9}{10.8}{\rmdefault}{\mddefault}{\updefault}{\color[rgb]{0,0,0}$(j=1, \dots, N )$}%
}}}}
\put(2997,-14820){\makebox(0,0)[lb]{\smash{{\SetFigFont{9}{10.8}{\rmdefault}{\mddefault}{\updefault}{\color[rgb]{0,0,0}$r_0$}%
}}}}
\put(591,-14834){\makebox(0,0)[lb]{\smash{{\SetFigFont{9}{10.8}{\rmdefault}{\mddefault}{\updefault}{\color[rgb]{0,0,0}$r_0$}%
}}}}
\put(5633,-15019){\makebox(0,0)[lb]{\smash{{\SetFigFont{9}{10.8}{\rmdefault}{\mddefault}{\updefault}{\color[rgb]{0,0,0}$r_0$}%
}}}}
%  0 
\put(3478,-14523){\makebox(0,0)[lb]{\smash{{\SetFigFont{9}{10.8}{\rmdefault}{\mddefault}{\updefault}{\color[rgb]{0,0,0}$z_{1+3i}$}%
}}}}
%  0 
\put(4876,-14386){\makebox(0,0)[lb]{\smash{{\SetFigFont{9}{10.8}{\rmdefault}{\mddefault}{\updefault}{\color[rgb]{0,0,0}$\Omega$}%
}}}}
%  10 
\put(226,-14386){\makebox(0,0)[lb]{\smash{{\SetFigFont{9}{10.8}{\rmdefault}{\mddefault}{\updefault}{\color[rgb]{0,0,0}$\Omega$}%
}}}}
%  0 
\put(2626,-14386){\makebox(0,0)[lb]{\smash{{\SetFigFont{9}{10.8}{\rmdefault}{\mddefault}{\updefault}{\color[rgb]{0,0,0}$\Omega$}%
}}}}
\end{picture}%
\caption{\raggedright Schematic view of (a) in-phase
  synchronization ($M=1$), (b) a three-cluster ($M=3$), and (c) a splay state
($M=N$). Each cluster consists of the same number of nodes. \label{fig:schema}}
\end{figure}
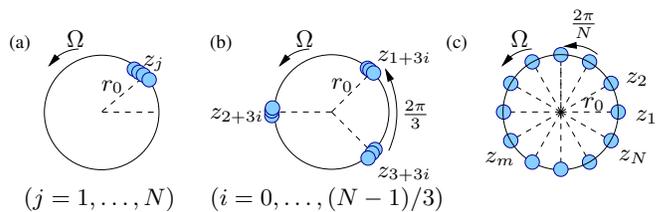

Cluster states are possible solutions
of Eq.~\eqref{eq:sl_r}. They exhibit a common 
amplitude $r_j\equiv r_{0}$ and phases given by $\varphi_j=\Omega_m
t+j 2\pi /M$, where $\Omega$ is  the collective frequency. The
integer $M$ determines the number of clusters. Here, we assume that
$M$ is a factor of $N$. Special clusters states
are in-phase  (also called zero-lag) synchronization ($M=1$), where all nodes are in one cluster, and the splay state
($M=N$), where each cluster consists of a single node. In the continuum limit, the
splay state on a unidirectionally coupled ring corresponds to a rotating wave. For a
schematic diagram of (a) in-phase synchronization, (b) a 3-cluster
state, and (c) a splay state see Fig.~\ref{fig:schema}

The stability of synchronized oscillations in networks can be
determined numerically, for instance, by the \textit{master stability
function}~\cite{PEC98}. This formalism allows for a separation of the
local dynamics of the individual nodes from the network topology. In
the case of the Stuart-Landau oscillators it is possible to obtain
the Floquet exponents of different cluster states analytically with
this technique \cite{CHO09}. By these means it has been demonstrated
that the unidirectional ring configuration, i.e.,
$G_{ij}=\delta_{(i+1) \bmod N,j}$, of Stuart-Landau
oscillators exhibits multistability  between in-phase synchrony, splay states, and clustering
in a large parameter range. The main goal of the adaptive topology developed
here is to suppress the multistability and to stabilize
the desired cluster state even for parameters for which it would be
unstable in the unidirectional ring. To find an appropriate adaptive
control, we make use of the speed gradient method \cite{FRA07}, which
is outlined in the next section.

Note that in theory synchronized
states could coexist, where the phase lags between different clusters are
of unequal size \cite{BLA13}. Such a phenomenon is called phase multistability and
could be an additional challenge to our control method. However, such
states have not been observed in the Stuart-Landau oscillator probably
because of the sinusoidal form of its oscillations. Furthermore, our
goal function (see next Section) is designed such that it has in
general only a minimum for the desired cluster states and thus would
prefer this state over states with unequal phase lags.

%%%%%%%%%%%%%%%%%%%%%%%%%%%%%%%%%%%%%%%%%%%%%%%%%%%%%%%%%%%%%%%%%%%%%%
\section{Speed-gradient method and goal function } \label{sec:sgm}
The main ingredient of the speed-gradient method \cite{FRA07} is a goal
function $Q(z(t),t)$ that is zero for the target state, i.e., a state
consisting of $M$ equally sized clusters with a constant phase lag 
% of $2\pi/d$
between subsequent clusters, and larger than zero otherwise.

An appropriate goal function $\tilde Q_M$ to control an $M$-cluster state was
suggested in Refs.~\cite{SEL12,SCH12}: 
\begin{eqnarray}
 \tilde  Q_M&=&1-f_M(\varphi)+\frac{N^2}{2}\sum_{p|M, 1\le
    p<M}f_p(\varphi),
\label{goalPhase4sel}
\end{eqnarray}
where $p|M$ denotes  that $p$ is a factor of $M$.
%The term $q$ is identical to the goal function $Q_4$
%in\cite{SEL12,SCH12}.
The function $f_M(\varphi)=\frac{1}{N^2}\sum_{j=1}^N
e^{iM\varphi_j}\sum_{k=1}^N
e^{-iM\varphi_k}$, where $\varphi_j$ is the phase of the $j$th
oscillator, $j=1,\ldots,N$, and $\varphi \equiv (\varphi_1,\ldots,\varphi_N)$, is a generalization of the Kuramoto order
parameter \cite{KUR84} and approaches unity for an $M$-cluster state.
% \begin{equation}
% f_M(\varphi)=\frac{1}{N^2}\sum_{j=1}^N e^{pi\varphi_j}\sum_{k=1}^N
% e^{-pi\varphi_k}, 
% \label{def_f_p}
% \end{equation} where $\varphi_j$ is the phase of the $j$th oscillator.
Since $f_M=1$ not only holds in the desired $M$-cluster state, but
also for all divisors $p$ of $M$, a goal function of the form
$\tilde Q_M=1-f_M(\varphi)$ would also vanish if the system were in one of
the  $p$-states. Therefore, the term $\sum_{p|M,
1\le p<M}f_p(\varphi)$ in
Eq.~\eqref{goalPhase4sel} was added as a penalty
term. 

Equation~\eqref{goalPhase4sel} can be used to control cluster
synchronization by means of adapting the topology and, in fact, the results are
qualitatively similar to the ones presented in this paper. However,
Eq.~\eqref{goalPhase4sel} has two drawbacks: First, it only ensures phase
synchronization, i.e., the radii of the oscillators do
not synchronize, and, second, the clusters are not of equal size. Therefore, we
use an extended version of Eq.~\eqref{goalPhase4sel}:
\begin{eqnarray}
  Q_M&=&\underbrace{1-f_M(\varphi)+\frac{N^2}{2}\sum_{ 1\le
    p<M}f_p(\varphi)+\frac{1}{2}\sum_{i,k}(r_i-r_k)^2}_{q_M} \nonumber \\&+&\frac{c}{2}\int_0^t \sum_k
\left(\sum_i G_{ki}(t')-1\right)^2dt'.
\label{goalPhase4}
\end{eqnarray}
We drop the condition $p|M$, in other words we add penalty terms for
all $p$-cluster states, where $p<M$. This ensures clusters of equal
size since it prevents  side maxima of the goal
function. Synchronization of the radii is reached due to the term
$\sum_{i,k}(r_i-r_k)^2$, where $r_i$ is the amplitude of the $i$th
oscillator. The last term in Eq.~\eqref{goalPhase4}, where $(G_{ki})$ is
the $N \times N$ coupling matrix and $c>0$ is a parameter, yields a unity row
sum. Without it, a constant but arbitrary row sum would
arise. Ensuring unity row sum helps to avoid side effects of changing
the effective coupling strengths by this arbitrary row sum.
% Thus, the term helps to separate the effects of changing topology
%from the adaptation of the effective coupling strength. 
As this term takes into account all deviations from the
unity-row sum during the growth of the network, $Q_M$ will not
vanish completely in the goal state. Thus, $q_M$ is a better measure
for the quality of synchronization. Though this might  be regarded as
a disadvantage of the integral in the
unity-row-sum term, the advantage of the integral is
that only terms in $G_{ij}$ but not in  $\dot G_{ij}$  appear in the right hand
side of the speed-gradient algorithm. Thus, it is not necessary to
solve for $\dot G_{ij}$.

The speed-gradient method \cite{FRA07} is generally used in the control of nonlinear
systems of the form $\dot z = F(z, u, t)$, where
$z\in\mathbb{C}^n$ is the state vector, $u\in\mathbb{C}^m$ is an input
(control) variable, and $F$ is a nonlinear function. The speed (or rate) at
which $Q(z(t),t)$ is changing along the trajectory of $z(t,u)$ is given by
$v(z,u,t)=\dot Q=\partial Q(z(t),t)/\partial t+[\nabla_z Q(z(t),t)]^T F(z,u,t)$.
The idea of the speed-gradient method is to change the control $u$ in the
direction of the negative gradient of $v(z,u,t)$ with respect to
the input variables: 
\begin{equation}
\label{speed-gradient algorithm} \frac{du}{dt}=-\gamma\nabla_uv(z,u,t),
\end{equation}
where $\gamma$ is a positive definite gain
matrix. The intuition behind the speed-gradient method is that $\dot Q$  may decrease
along the direction of its negative gradient and as $\dot  Q$  becomes negative, 
$Q$  will decrease as well and may finally tend to zero, i.e., the control goal is reached.
In fact those statements hold under some additional conditions \cite{AND99a,FRA96}.

%%%%%%%%%%%%%%%%%%%%%%%%%%%%%%%%%%%%%%%%%%%%%%%%%%%%%%%%%%%%%%%%%%%%%%%%%%%%%%%%%%%%%%%%%%%%
\section{ Adapting the topology } \label{sec:adapting-topology-}
\begin{widetext}

\begin{figure}[ht]
\raggedright
\captionsetup[subfigure]{labelformat=empty, justification=raggedleft}
 %\captionsetup[subfigure]{labelfont=bf,textfont=normalfont,singlelinecheck=off,justification=raggedright}
 \subfloat[(a) \:$t=-50$ (initial state)]{
\adjustbox{trim={.15\width} {.15\height} {0.15\width} {.15\height},clip}{
  \includegraphics[width=.30\linewidth,
  height=.30\linewidth]{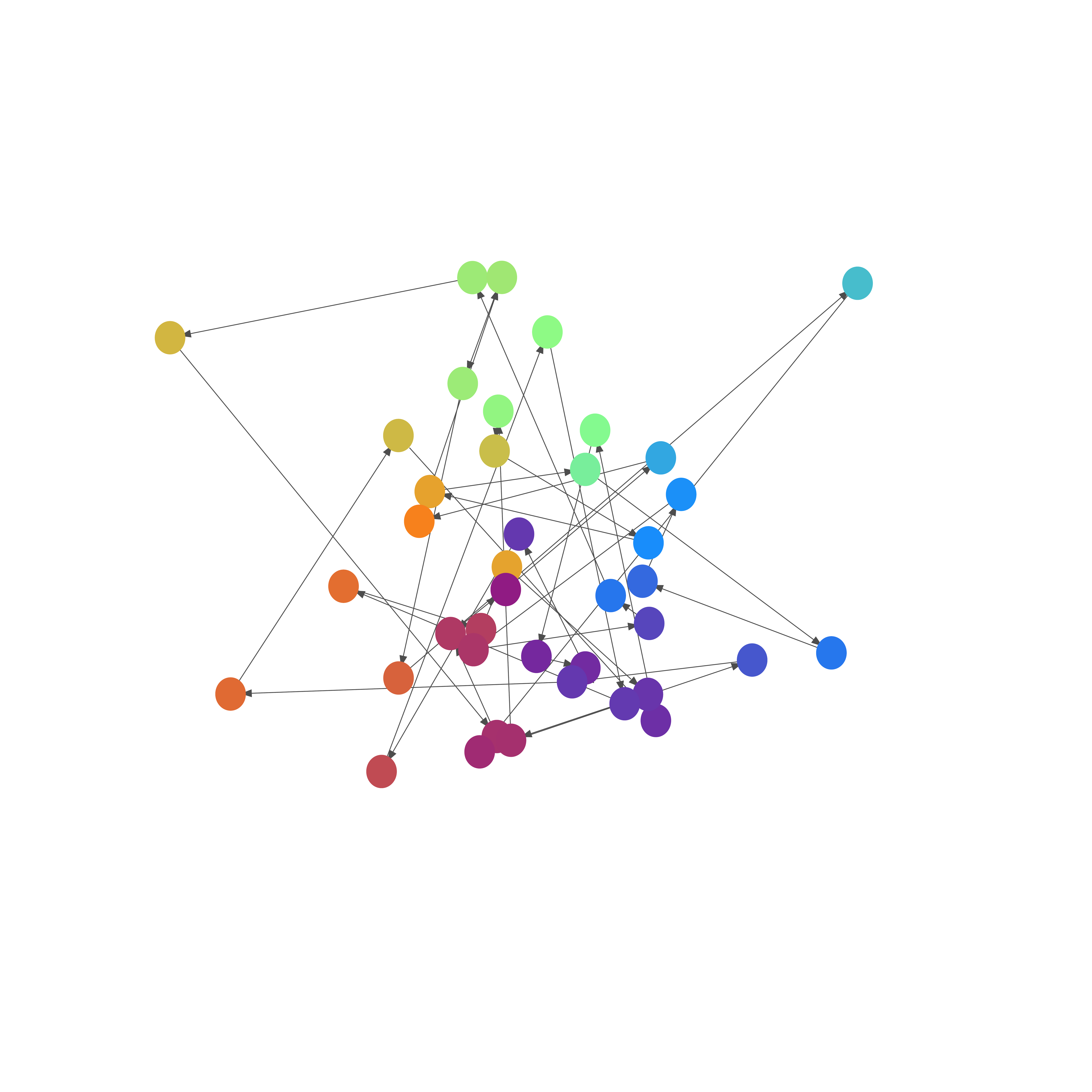}}}
\subfloat[(b) \:$t=0$ (control starts)]{
\adjustbox{trim={.15\width} {.15\height} {0.15\width} {.15\height},clip}{
  \includegraphics[width=.30\linewidth,
  height=.30\linewidth]{./controlstart.pdf}}}
\subfloat[(c) \:$t=2$ (shortly after control started)]{
\adjustbox{trim={.15\width} {.15\height} {0.15\width} {.15\height},clip}{
  \includegraphics[width=.30\linewidth,
  height=.30\linewidth]{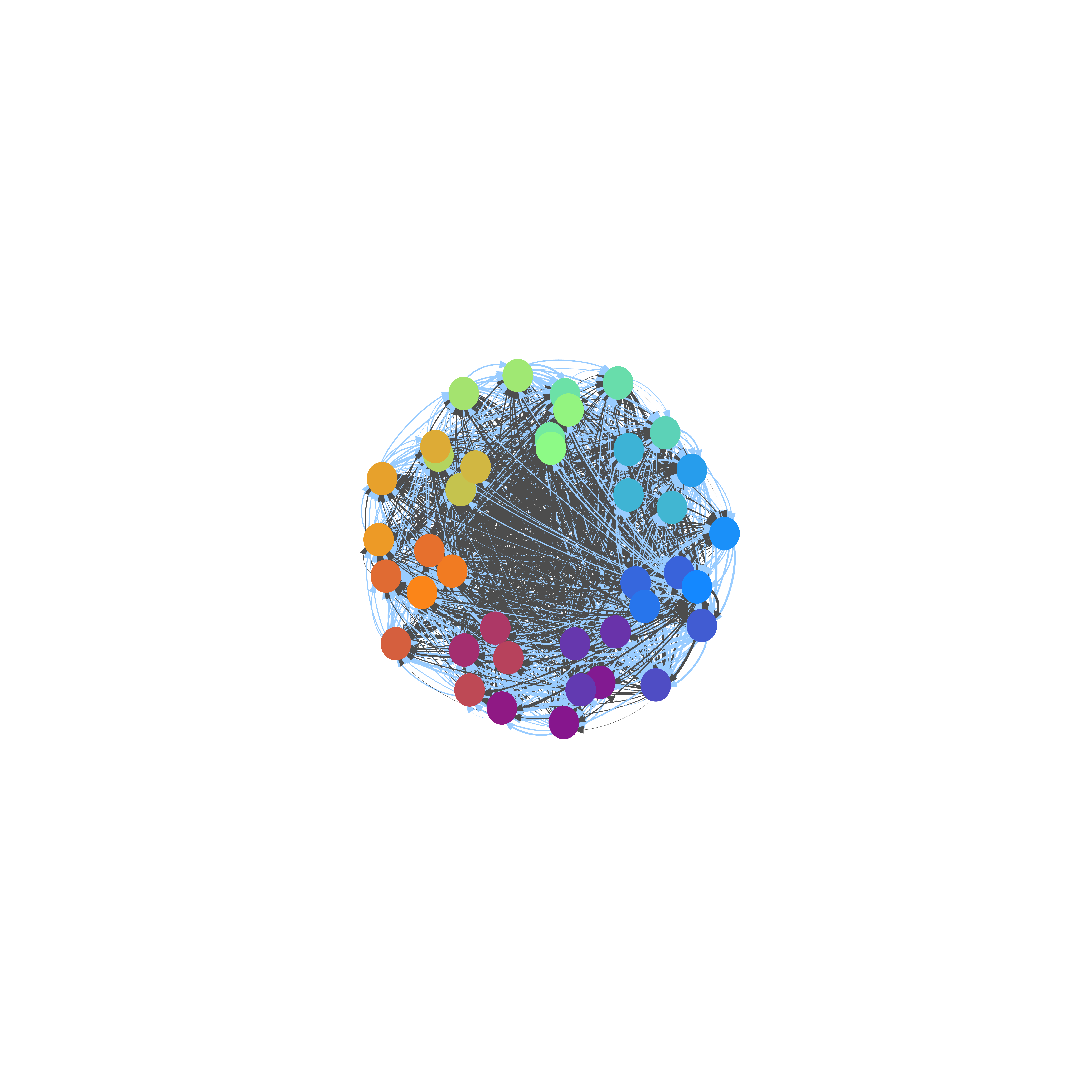}}}
\subfloat[(d) \:$t=50$ (final state)]{
\adjustbox{trim={.15\width} {.15\height} {0.15\width} {.15\height},clip}{
  \includegraphics[width=.30\linewidth,
  height=.30\linewidth]{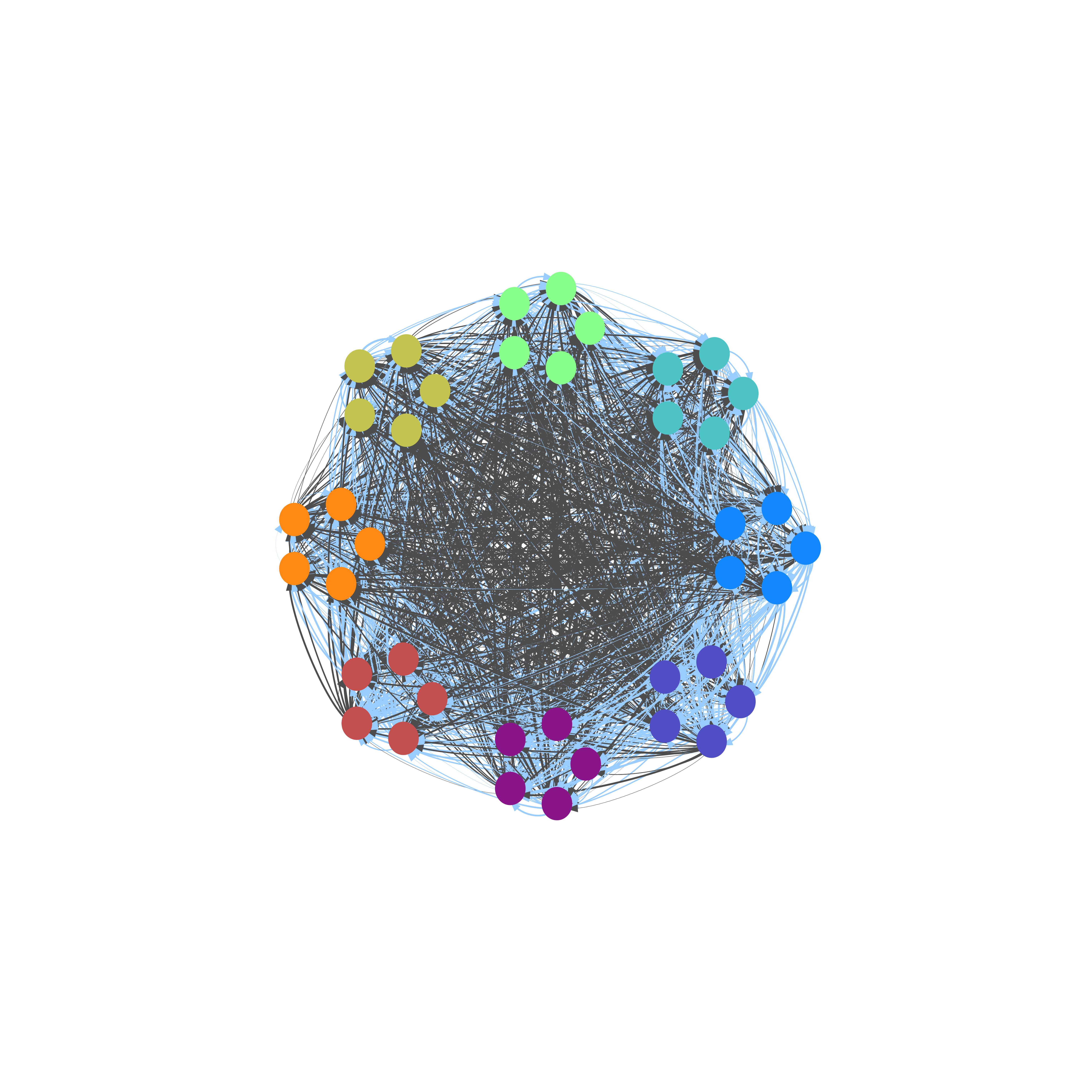}}}
 \subfloat{
 \raisebox{-.70cm}{\adjustbox{trim={.2\width} {.15\height} {0.25\width} {.1\height},clip}{
  \includegraphics[width=.2\linewidth,
  height=.40\linewidth]{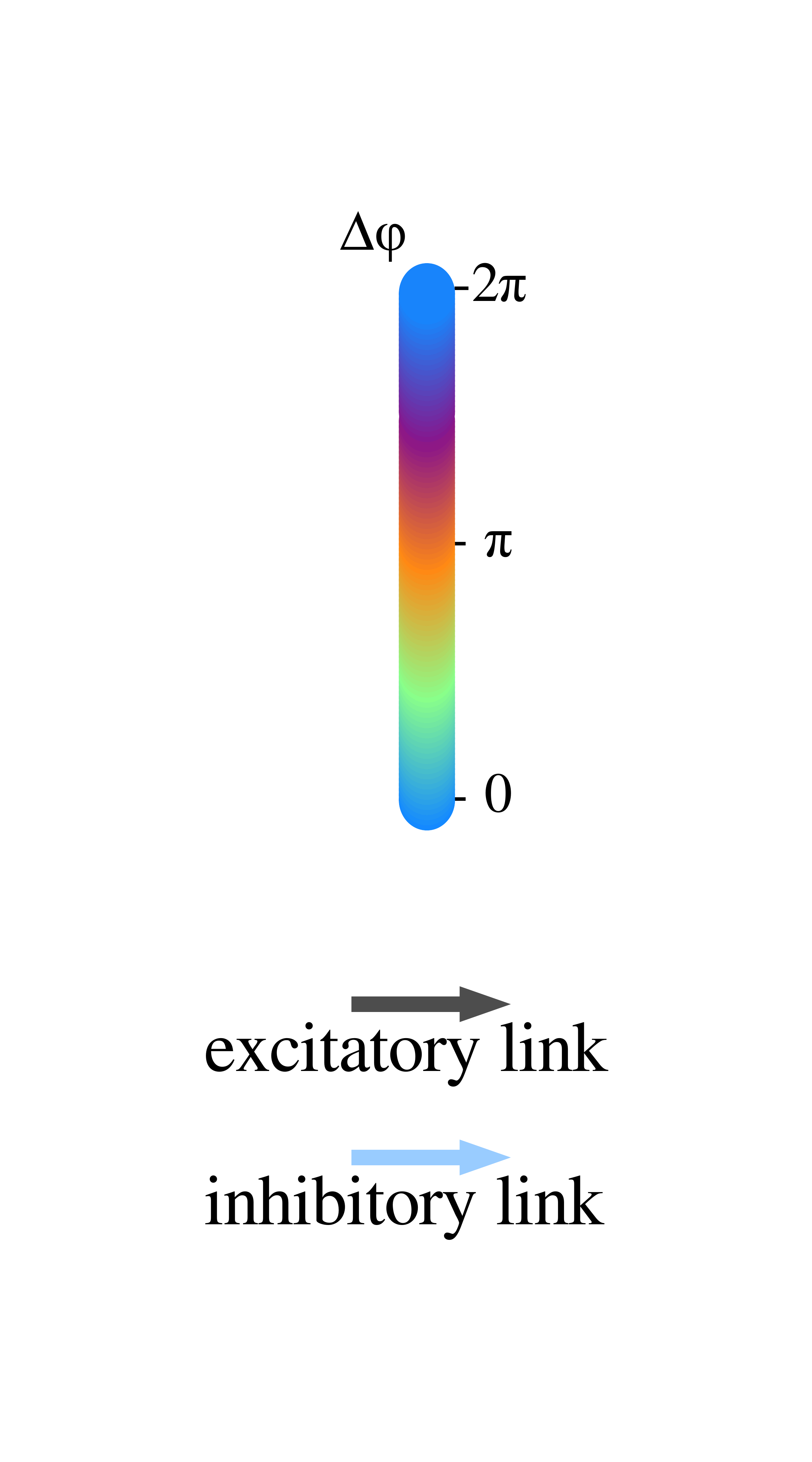}}}}
% trim=l b r t
\vspace{-0.2cm}
\caption[]{
\raggedright Evolution of the  network topology with the goal to achieve an 8-cluster state. Black:
  excitatory weighted links; gray (blue)  inhibitory
  weighted links. Node colors denote phase differences with respect to
the first node. Parameters: $\lambda=0.1$, $\omega=1$, $c=0.01$, $K=0.1$, $\tau=\pi$, $\gamma_G=10$, $N=40$, $M=8$.
 \label{fig:network8}

}
\end{figure}
\end{widetext}

\begin{figure}[]  

% GNUPLOT: LaTeX picture with Postscript
\begingroup
  \makeatletter
  \providecommand\color[2][]{%
    \GenericError{(gnuplot) \space\space\space\@spaces}{%
      Package color not loaded in conjunction with
      terminal option `colourtext'%
    }{See the gnuplot documentation for explanation.%
    }{Either use 'blacktext' in gnuplot or load the package
      color.sty in LaTeX.}%
    \renewcommand\color[2][]{}%
  }%
  \providecommand\includegraphics[2][]{%
    \GenericError{(gnuplot) \space\space\space\@spaces}{%
      Package graphicx or graphics not loaded%
    }{See the gnuplot documentation for explanation.%
    }{The gnuplot epslatex terminal needs graphicx.sty or graphics.sty.}%
    \renewcommand\includegraphics[2][]{}%
  }%
  \providecommand\rotatebox[2]{#2}%
  \@ifundefined{ifGPcolor}{%
    \newif\ifGPcolor
    \GPcolortrue
  }{}%
  \@ifundefined{ifGPblacktext}{%
    \newif\ifGPblacktext
    \GPblacktexttrue
  }{}%
  % define a \g@addto@macro without @ in the name:
  \let\gplgaddtomacro\g@addto@macro
  % define empty templates for all commands taking text:
  \gdef\gplbacktext{}%
  \gdef\gplfronttext{}%
  \makeatother
  \ifGPblacktext
    % no textcolor at all
    \def\colorrgb#1{}%
    \def\colorgray#1{}%
  \else
    % gray or color?
    \ifGPcolor
      \def\colorrgb#1{\color[rgb]{#1}}%
      \def\colorgray#1{\color[gray]{#1}}%
      \expandafter\def\csname LTw\endcsname{\color{white}}%
      \expandafter\def\csname LTb\endcsname{\color{black}}%
      \expandafter\def\csname LTa\endcsname{\color{black}}%
      \expandafter\def\csname LT0\endcsname{\color[rgb]{1,0,0}}%
      \expandafter\def\csname LT1\endcsname{\color[rgb]{0,1,0}}%
      \expandafter\def\csname LT2\endcsname{\color[rgb]{0,0,1}}%
      \expandafter\def\csname LT3\endcsname{\color[rgb]{1,0,1}}%
      \expandafter\def\csname LT4\endcsname{\color[rgb]{0,1,1}}%
      \expandafter\def\csname LT5\endcsname{\color[rgb]{1,1,0}}%
      \expandafter\def\csname LT6\endcsname{\color[rgb]{0,0,0}}%
      \expandafter\def\csname LT7\endcsname{\color[rgb]{1,0.3,0}}%
      \expandafter\def\csname LT8\endcsname{\color[rgb]{0.5,0.5,0.5}}%
    \else
      % gray
      \def\colorrgb#1{\color{black}}%
      \def\colorgray#1{\color[gray]{#1}}%
      \expandafter\def\csname LTw\endcsname{\color{white}}%
      \expandafter\def\csname LTb\endcsname{\color{black}}%
      \expandafter\def\csname LTa\endcsname{\color{black}}%
      \expandafter\def\csname LT0\endcsname{\color{black}}%
      \expandafter\def\csname LT1\endcsname{\color{black}}%
      \expandafter\def\csname LT2\endcsname{\color{black}}%
      \expandafter\def\csname LT3\endcsname{\color{black}}%
      \expandafter\def\csname LT4\endcsname{\color{black}}%
      \expandafter\def\csname LT5\endcsname{\color{black}}%
      \expandafter\def\csname LT6\endcsname{\color{black}}%
      \expandafter\def\csname LT7\endcsname{\color{black}}%
      \expandafter\def\csname LT8\endcsname{\color{black}}%
    \fi
  \fi
  \setlength{\unitlength}{0.0500bp}%
  \begin{picture}(4860.00,3400.00)%
    \gplgaddtomacro\gplbacktext{%
      \csname LTb\endcsname%
      \put(369,1793){\makebox(0,0)[r]{\strut{} 0}}%
      \csname LTb\endcsname%
      \put(369,2055){\makebox(0,0)[r]{\strut{} 1}}%
      \csname LTb\endcsname%
      \put(369,2317){\makebox(0,0)[r]{\strut{} 2}}%
      \csname LTb\endcsname%
      \put(369,2579){\makebox(0,0)[r]{\strut{} 3}}%
      \csname LTb\endcsname%
      \put(369,2841){\makebox(0,0)[r]{\strut{} 4}}%
      \csname LTb\endcsname%
      \put(587,1607){\makebox(0,0){\strut{} }}%
      \csname LTb\endcsname%
      \put(819,1607){\makebox(0,0){\strut{} }}%
      \csname LTb\endcsname%
      \put(1052,1607){\makebox(0,0){\strut{} }}%
      \csname LTb\endcsname%
      \put(1284,1607){\makebox(0,0){\strut{} }}%
      \csname LTb\endcsname%
      \put(1516,1607){\makebox(0,0){\strut{} }}%
      \csname LTb\endcsname%
      \put(1748,1607){\makebox(0,0){\strut{} }}%
      \csname LTb\endcsname%
      \put(1981,1607){\makebox(0,0){\strut{} }}%
      \csname LTb\endcsname%
      \put(2213,1607){\makebox(0,0){\strut{} }}%
      \csname LTb\endcsname%
      \put(72,2317){\rotatebox{-270}{\makebox(0,0){\strut{}$r_j$}}}%
    }%
    \gplgaddtomacro\gplfronttext{%
      \csname LTb\endcsname%
      \put(2143,2684){\makebox(0,0){\strut{}(a)}}%
    }%
    \gplgaddtomacro\gplbacktext{%
      \csname LTb\endcsname%
      \put(369,558){\makebox(0,0)[r]{\strut{}0}}%
      \csname LTb\endcsname%
      \put(369,1083){\makebox(0,0)[r]{\strut{}$\pi$}}%
      \csname LTb\endcsname%
      \put(369,1607){\makebox(0,0)[r]{\strut{}$2\pi$}}%
      \csname LTb\endcsname%
      \put(471,372){\makebox(0,0){\strut{}-50}}%
      \csname LTb\endcsname%
      \put(1052,372){\makebox(0,0){\strut{}0}}%
      \csname LTb\endcsname%
      \put(1632,372){\makebox(0,0){\strut{}50}}%
      \csname LTb\endcsname%
      \put(2213,372){\makebox(0,0){\strut{}100}}%
      \csname LTb\endcsname%
      \put(72,1082){\rotatebox{-270}{\makebox(0,0){\strut{}$\varphi_j-\varphi_0$}}}%
      \csname LTb\endcsname%
      \put(1400,93){\makebox(0,0){\strut{}t}}%
    }%
    \gplgaddtomacro\gplfronttext{%
      \csname LTb\endcsname%
      \put(2143,1450){\makebox(0,0){\strut{}(b)}}%
    }%
    \gplgaddtomacro\gplbacktext{%
      \csname LTb\endcsname%
      \put(2847,1943){\makebox(0,0)[r]{\strut{}-5}}%
      \csname LTb\endcsname%
      \put(2847,2317){\makebox(0,0)[r]{\strut{}0}}%
      \csname LTb\endcsname%
      \put(2847,2691){\makebox(0,0)[r]{\strut{}5}}%
      \csname LTb\endcsname%
      \put(2949,1607){\makebox(0,0){\strut{} }}%
      \csname LTb\endcsname%
      \put(3530,1607){\makebox(0,0){\strut{} }}%
      \csname LTb\endcsname%
      \put(4111,1607){\makebox(0,0){\strut{} }}%
      \csname LTb\endcsname%
      \put(4692,1607){\makebox(0,0){\strut{} }}%
      \csname LTb\endcsname%
      \put(2550,2317){\rotatebox{-270}{\makebox(0,0){\strut{}$G_{jn}$}}}%
    }%
    \gplgaddtomacro\gplfronttext{%
      \csname LTb\endcsname%
      \put(4622,2684){\makebox(0,0){\strut{}(c)}}%
    }%
    \gplgaddtomacro\gplbacktext{%
      \csname LTb\endcsname%
      \put(2847,558){\makebox(0,0)[r]{\strut{}0}}%
      \csname LTb\endcsname%
      \put(2847,1064){\makebox(0,0)[r]{\strut{}20}}%
      \csname LTb\endcsname%
      \put(2847,1570){\makebox(0,0)[r]{\strut{}40}}%
      \csname LTb\endcsname%
      \put(2949,372){\makebox(0,0){\strut{}-50}}%
      \csname LTb\endcsname%
      \put(3530,372){\makebox(0,0){\strut{}0}}%
      \csname LTb\endcsname%
      \put(4111,372){\makebox(0,0){\strut{}50}}%
      \csname LTb\endcsname%
      \put(4692,372){\makebox(0,0){\strut{}100}}%
      \csname LTb\endcsname%
      \put(2550,1064){\rotatebox{-270}{\makebox(0,0){\strut{}$Q_8,q_8$}}}%
      \csname LTb\endcsname%
      \put(3878,93){\makebox(0,0){\strut{}t}}%
    }%
    \gplgaddtomacro\gplfronttext{%
      \csname LTb\endcsname%
      \put(4056,1376){\makebox(0,0)[r]{\strut{}$Q_8$}}%
      \csname LTb\endcsname%
      \put(4056,1190){\makebox(0,0)[r]{\strut{}$q_8$}}%
      \csname LTb\endcsname%
      \put(4622,1418){\makebox(0,0){\strut{}(d)}}%
    }%
    \gplbacktext
    \put(0,0){\includegraphics{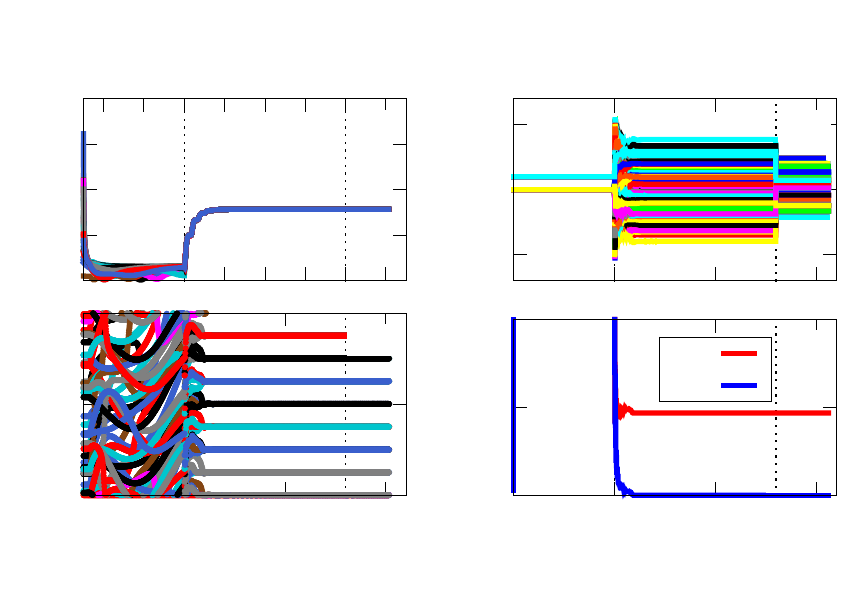}}%
    \gplfronttext
  \end{picture}%
\endgroup
%\input{./zeitserieneu.tex}
%~/topology/koeffizienten $ 
\vspace{-0.2cm}
  \caption{ \raggedright
 Control of an 8-cluster state: (a) radii $r_j$, (b) phase
    difference $\varphi_j-\varphi_0$ with respect to
    the first node, (c) coupling weights $G_{ij}$, and (d) goal
    function $Q_8$ and its reduced part $q_8$ (excluding the
    unity-row-sum term). Control starts at $t=0$. The vertical dotted
    line at $t=80$ is explained in Sec.~\ref{sec:stab-clust-stat}. Parameters as in Fig.~\ref{fig:network8}.}
  \label{fig:zeitserie1}
\end{figure}

Using Eq.~\eqref{speed-gradient algorithm} with $u=(G_{11}, G_{12},\ldots ,G_{N,N-1},G_{NN}) \in
\mathbb{R}^{N^2}$ and $\gamma_{ij}= \delta_{ij}\gamma_G$, where
$\gamma_G$ is a positive constant, we obtain
the following control scheme:
\begin{widetext}
\begin{eqnarray} \dot{G}_{jn}(t)&=& -\gamma_G K
\left[\frac{r_{n,\tau}}{r_j}\sin(\varphi_{n,\tau}
-\varphi_j)\right] \times
\sum_{k=1}^N\left\{\sum_{1\le
p<M}p\sin[p(\varphi_k-\varphi_j)]-\frac{2M}{N^2}
\sin[M(\varphi_k-\varphi_j)]\right\}
\nonumber \\ &&
-2\gamma_G K\sum_k(r_j-r_k)\left[r_{n,\tau}\cos(\varphi_{n,\tau}
-\varphi_j)-r_j\right]-\gamma_G c(\sum_i G_{ji} -1).
\label{controlalgo}
\end{eqnarray}
\end{widetext}
% where we used polar coordinates for $z$:
%$z_j=r_je^{i\varphi_j}$.

Figure~\ref{fig:network8} shows an example of the evolution of the network by
applying the control algorithm~\eqref{controlalgo} with $Q_8$, i.e.,
the goal to reach an 8-cluster state. The
initial topology is a unidirectional ring, see
Fig.~\ref{fig:network8}(a). However, the nodes are approximately
depicted according to their position $z_j$ in the phase space; because these phase differences are initially
random, the unidirectional coupling structure does not clearly
 show here. At $t=0$ the control is switched on (Fig.~\ref{fig:network8}(b)), and links change
rapidly as can be seen in Fig.~\ref{fig:network8}(c). The final state
of the network is shown Fig.~\ref{fig:network8}(d): 8 equally
sized clusters are formed (to distinguish all nodes, the nodes in one cluster are
not depicted exactly according to their phases but on a circle around
the point which would correspond to their phase and radius). Black links mark
excitatory links, i.e., links that correspond to positive entries of
the coupling matrix, while the gray (light blue) links are inhibitory
ones, i.e., links due to negative entries of the coupling matrix.
Clearly, the final topology is characterized by a distinct
distribution of excitatory and inhibitory links: While the inhibitory
links mainly connect neighboring clusters, the excitatory ones
dominate the connections to clusters further away in phase space. In
the next Section this distribution is investigated more closely. 
 
Corresponding to the network realization in
Fig.~\ref{fig:network8}, Fig.~\ref{fig:zeitserie1} shows the time series of (a) the radii,
(b) the phase differences, (c) the coupling weights, and (d) the goal
function. After the control is switched on, the radii and phases
rapidly converge to the 8-cluster state, and $Q_8$ and $q_8$ approach
their minimum.

\section{ Structural properties } \label{sec:struct-prop-}

\begin{figure}[]
\centering
% GNUPLOT: LaTeX picture with Postscript
\begingroup
  \makeatletter
  \providecommand\color[2][]{%
    \GenericError{(gnuplot) \space\space\space\@spaces}{%
      Package color not loaded in conjunction with
      terminal option `colourtext'%
    }{See the gnuplot documentation for explanation.%
    }{Either use 'blacktext' in gnuplot or load the package
      color.sty in LaTeX.}%
    \renewcommand\color[2][]{}%
  }%
  \providecommand\includegraphics[2][]{%
    \GenericError{(gnuplot) \space\space\space\@spaces}{%
      Package graphicx or graphics not loaded%
    }{See the gnuplot documentation for explanation.%
    }{The gnuplot epslatex terminal needs graphicx.sty or graphics.sty.}%
    \renewcommand\includegraphics[2][]{}%
  }%
  \providecommand\rotatebox[2]{#2}%
  \@ifundefined{ifGPcolor}{%
    \newif\ifGPcolor
    \GPcolortrue
  }{}%
  \@ifundefined{ifGPblacktext}{%
    \newif\ifGPblacktext
    \GPblacktexttrue
  }{}%
  % define a \g@addto@macro without @ in the name:
  \let\gplgaddtomacro\g@addto@macro
  % define empty templates for all commands taking text:
  \gdef\gplbacktext{}%
  \gdef\gplfronttext{}%
  \makeatother
  \ifGPblacktext
    % no textcolor at all
    \def\colorrgb#1{}%
    \def\colorgray#1{}%
  \else
    % gray or color?
    \ifGPcolor
      \def\colorrgb#1{\color[rgb]{#1}}%
      \def\colorgray#1{\color[gray]{#1}}%
      \expandafter\def\csname LTw\endcsname{\color{white}}%
      \expandafter\def\csname LTb\endcsname{\color{black}}%
      \expandafter\def\csname LTa\endcsname{\color{black}}%
      \expandafter\def\csname LT0\endcsname{\color[rgb]{1,0,0}}%
      \expandafter\def\csname LT1\endcsname{\color[rgb]{0,1,0}}%
      \expandafter\def\csname LT2\endcsname{\color[rgb]{0,0,1}}%
      \expandafter\def\csname LT3\endcsname{\color[rgb]{1,0,1}}%
      \expandafter\def\csname LT4\endcsname{\color[rgb]{0,1,1}}%
      \expandafter\def\csname LT5\endcsname{\color[rgb]{1,1,0}}%
      \expandafter\def\csname LT6\endcsname{\color[rgb]{0,0,0}}%
      \expandafter\def\csname LT7\endcsname{\color[rgb]{1,0.3,0}}%
      \expandafter\def\csname LT8\endcsname{\color[rgb]{0.5,0.5,0.5}}%
    \else
      % gray
      \def\colorrgb#1{\color{black}}%
      \def\colorgray#1{\color[gray]{#1}}%
      \expandafter\def\csname LTw\endcsname{\color{white}}%
      \expandafter\def\csname LTb\endcsname{\color{black}}%
      \expandafter\def\csname LTa\endcsname{\color{black}}%
      \expandafter\def\csname LT0\endcsname{\color{black}}%
      \expandafter\def\csname LT1\endcsname{\color{black}}%
      \expandafter\def\csname LT2\endcsname{\color{black}}%
      \expandafter\def\csname LT3\endcsname{\color{black}}%
      \expandafter\def\csname LT4\endcsname{\color{black}}%
      \expandafter\def\csname LT5\endcsname{\color{black}}%
      \expandafter\def\csname LT6\endcsname{\color{black}}%
      \expandafter\def\csname LT7\endcsname{\color{black}}%
      \expandafter\def\csname LT8\endcsname{\color{black}}%
    \fi
  \fi
  \setlength{\unitlength}{0.0500bp}%
  \begin{picture}(5832.00,3999.60)%
    \gplgaddtomacro\gplbacktext{%
      \csname LTb\endcsname%
      \put(543,3192){\makebox(0,0)[r]{\strut{}-2}}%
      \csname LTb\endcsname%
      \put(543,3445){\makebox(0,0)[r]{\strut{} 0}}%
      \csname LTb\endcsname%
      \put(543,3697){\makebox(0,0)[r]{\strut{} 2}}%
      \csname LTb\endcsname%
      \put(674,2778){\makebox(0,0){\strut{} }}%
      \csname LTb\endcsname%
      \put(1454,2778){\makebox(0,0){\strut{} }}%
      \csname LTb\endcsname%
      \put(3015,2778){\makebox(0,0){\strut{} }}%
      \csname LTb\endcsname%
      \put(3795,2778){\makebox(0,0){\strut{} }}%
      \csname LTb\endcsname%
      \put(144,3444){\rotatebox{-270}{\makebox(0,0){\strut{}$\bar G(n\frac{2\pi}{M})$}}}%
    }%
    \gplgaddtomacro\gplfronttext{%
      \csname LTb\endcsname%
      \put(4510,3718){\makebox(0,0)[r]{\strut{}$ \tau=0$}}%
      \csname LTb\endcsname%
      \put(4510,3495){\makebox(0,0)[r]{\strut{}$ \tau= 2\pi $}}%
    }%
    \gplgaddtomacro\gplbacktext{%
      \csname LTb\endcsname%
      \put(543,2083){\makebox(0,0)[r]{\strut{}-2}}%
      \csname LTb\endcsname%
      \put(543,2336){\makebox(0,0)[r]{\strut{} 0}}%
      \csname LTb\endcsname%
      \put(543,2589){\makebox(0,0)[r]{\strut{} 2}}%
      \csname LTb\endcsname%
      \put(674,1670){\makebox(0,0){\strut{} }}%
      \csname LTb\endcsname%
      \put(1454,1670){\makebox(0,0){\strut{} }}%
      \csname LTb\endcsname%
      \put(3015,1670){\makebox(0,0){\strut{} }}%
      \csname LTb\endcsname%
      \put(3795,1670){\makebox(0,0){\strut{} }}%
      \csname LTb\endcsname%
      \put(144,2336){\rotatebox{-270}{\makebox(0,0){\strut{}$\bar G(n\frac{2\pi}{M})$}}}%
    }%
    \gplgaddtomacro\gplfronttext{%
      \csname LTb\endcsname%
      \put(4510,2609){\makebox(0,0)[r]{\strut{}$ \tau=\frac{\pi}{2} $}}%
      \csname LTb\endcsname%
      \put(4510,2386){\makebox(0,0)[r]{\strut{}$ \tau=\frac{3\pi}{2} $}}%
    }%
    \gplgaddtomacro\gplbacktext{%
      \csname LTb\endcsname%
      \put(543,972){\makebox(0,0)[r]{\strut{}-2}}%
      \csname LTb\endcsname%
      \put(543,1226){\makebox(0,0)[r]{\strut{} 0}}%
      \csname LTb\endcsname%
      \put(543,1480){\makebox(0,0)[r]{\strut{} 2}}%
      \csname LTb\endcsname%
      \put(674,558){\makebox(0,0){\strut{}$-\pi $}}%
      \csname LTb\endcsname%
      \put(1454,558){\makebox(0,0){\strut{}$-\frac{\pi}{2} $}}%
      \csname LTb\endcsname%
      \put(2235,558){\makebox(0,0){\strut{}0}}%
      \csname LTb\endcsname%
      \put(3015,558){\makebox(0,0){\strut{}$\frac{\pi}{2}$}}%
      \csname LTb\endcsname%
      \put(3795,558){\makebox(0,0){\strut{}$ \pi $}}%
      \csname LTb\endcsname%
      \put(144,1226){\rotatebox{-270}{\makebox(0,0){\strut{}$\bar G(n\frac{2\pi}{M})$}}}%
      \csname LTb\endcsname%
      \put(2234,279){\makebox(0,0){\strut{}$\Delta_{ij}$}}%
    }%
    \gplgaddtomacro\gplfronttext{%
      \csname LTb\endcsname%
      \put(4510,1501){\makebox(0,0)[r]{\strut{}$ \tau=\pi $}}%
      \csname LTb\endcsname%
      \put(4510,1278){\makebox(0,0)[r]{\strut{}$ \tau=3\pi $}}%
    }%
    \gplbacktext
    \put(0,0){\includegraphics{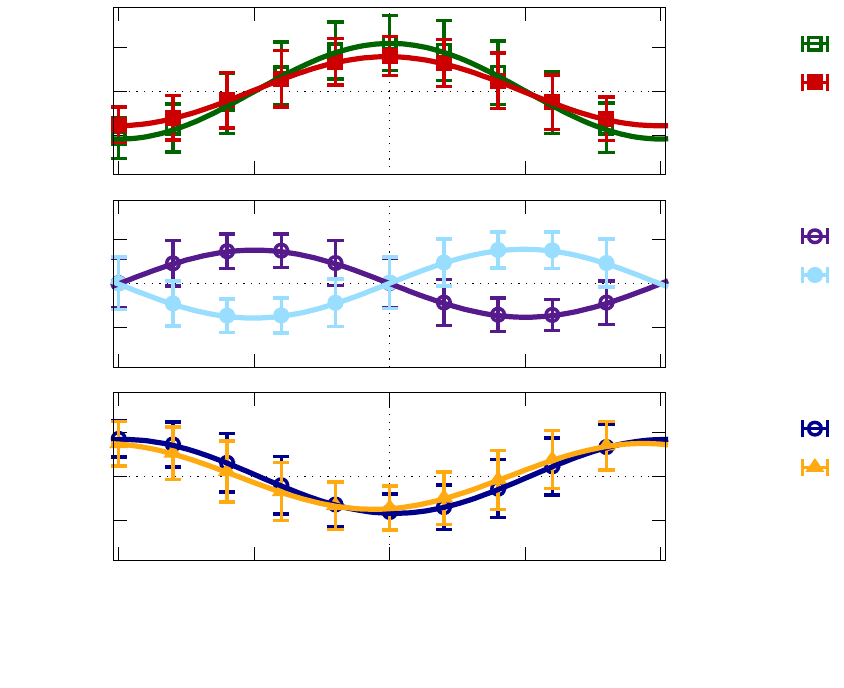}}%
    \gplfronttext
  \end{picture}%
\endgroup

\caption{\raggedright Average link strength $\bar G(\frac{2\pi}{M}
  n)$ versus phase differences $\Delta_{ij}=\lim_{t \rightarrow
    \infty}{[\varphi_i(t)-\varphi_j(t)]}$ in the 
final state as defined in  Eq.~\eqref{eq:def_G_bar}. 
% , i.e. in the interval $[2\pi j/d\pi/d,2\pi
% j/d+\pi/d]$, $j=(-d-1)/2,\ldots,(-d+1)/2$, for the shown realizations
% realizations. 
$N=30$, $M=10$. Average over 100 realizations. Other parameters as in Fig.~\ref{fig:network8}.
\label{weightvsd}}
%\end{minipage}  
\end{figure}

%In complex networks, a synchronized solution exists only in the case
%that the network fulfills certain conditions. In the most simple case
%of in-phase synchronization constant the condition of a constant row
%sum, i.e. $\sum_j G_{i,j}$, is sufficient for the existence a in-phase
%synchronized solution \cite{PEC98}. In the case of cluster
%synchronization, the topology underlies additional constraints. For
%example, multipartite    
In the field of complex networks, topological features enhancing or the
weakening synchronizability are of great interest
\cite{PEC98,CHA05,LEH11,KEA12}. Here, we discuss the structural
properties of the networks
after successful control, in order to elucidate the common features that
enable synchronization in a cluster
state. To do so we consider the coupling weights of the final topology
as a function of the final phase difference, i.e., the function
$\tilde G_{ij}(\Delta_{ij})$ where $\Delta_{ij}\equiv \lim_{t \rightarrow \infty}[{\varphi_i(t)-\varphi_j(t)}]$.
%and  $t_\infty$ is some
%asymptotic  time  after the disappearance of all transient effects.
In Fig.~\ref{weightvsd} we plot
\begin{eqnarray}\label{eq:def_G_bar}
\bar G\left(\frac{2\pi}{M} n\right) =  \Biggl\langle  \hspace{0cm} \sum_{\substack{ij\\(\Delta_{ij})
    \in I_n }} 
%\hspace{-1cm} 
G_{ij}(\Delta_{ij}) 
\Biggr \rangle
\end{eqnarray} with the interval $I_n=(\frac{2\pi n-\pi}{M},\frac{2\pi
  n+\pi}{M}]$, $n=-M/2,\ldots, M/2-1$ if $M$ is even and
$n=-(M-1)/2,\ldots, (M-1)/2$ if $M$ is odd. $\langle \cdot \rangle$ denotes the ensemble
average over 100 realizations, i.e., $\bar G(\frac{2\pi}{M} n)$ is the
average of all weights linking nodes which have a phase difference in
the interval $I_n$. In the case
of successful control these are the weights linking nodes in cluster
$i$ with nodes in cluster $(i+n) \bmod M$.  Figure~\ref{weightvsd} depicts the weights for different delay
times. Obviously, the curves have the form of a time shifted cosine:
i.e., $\bar G(\frac{2\pi}{M} n) \propto \cos(\frac{2\pi n}{M}-\tau)$.
This explains the topological structure we observe in
Fig.~ \ref{fig:network8}(c): Since $\tau=\pi$, we expect
a structure as  described by
Fig.~\ref{weightvsd}(c). Thus, a negative coupling between nodes with a
small phase difference and a positive coupling between nodes with a
phase equal or close to $\pi$.

\subsection{ Existence of cluster solutions } \label{sec:exist-clust-solut}
\begin{figure}[]  

% GNUPLOT: LaTeX picture with Postscript
\begingroup
  \makeatletter
  \providecommand\color[2][]{%
    \GenericError{(gnuplot) \space\space\space\@spaces}{%
      Package color not loaded in conjunction with
      terminal option `colourtext'%
    }{See the gnuplot documentation for explanation.%
    }{Either use 'blacktext' in gnuplot or load the package
      color.sty in LaTeX.}%
    \renewcommand\color[2][]{}%
  }%
  \providecommand\includegraphics[2][]{%
    \GenericError{(gnuplot) \space\space\space\@spaces}{%
      Package graphicx or graphics not loaded%
    }{See the gnuplot documentation for explanation.%
    }{The gnuplot epslatex terminal needs graphicx.sty or graphics.sty.}%
    \renewcommand\includegraphics[2][]{}%
  }%
  \providecommand\rotatebox[2]{#2}%
  \@ifundefined{ifGPcolor}{%
    \newif\ifGPcolor
    \GPcolortrue
  }{}%
  \@ifundefined{ifGPblacktext}{%
    \newif\ifGPblacktext
    \GPblacktexttrue
  }{}%
  % define a \g@addto@macro without @ in the name:
  \let\gplgaddtomacro\g@addto@macro
  % define empty templates for all commands taking text:
  \gdef\gplbacktext{}%
  \gdef\gplfronttext{}%
  \makeatother
  \ifGPblacktext
    % no textcolor at all
    \def\colorrgb#1{}%
    \def\colorgray#1{}%
  \else
    % gray or color?
    \ifGPcolor
      \def\colorrgb#1{\color[rgb]{#1}}%
      \def\colorgray#1{\color[gray]{#1}}%
      \expandafter\def\csname LTw\endcsname{\color{white}}%
      \expandafter\def\csname LTb\endcsname{\color{black}}%
      \expandafter\def\csname LTa\endcsname{\color{black}}%
      \expandafter\def\csname LT0\endcsname{\color[rgb]{1,0,0}}%
      \expandafter\def\csname LT1\endcsname{\color[rgb]{0,1,0}}%
      \expandafter\def\csname LT2\endcsname{\color[rgb]{0,0,1}}%
      \expandafter\def\csname LT3\endcsname{\color[rgb]{1,0,1}}%
      \expandafter\def\csname LT4\endcsname{\color[rgb]{0,1,1}}%
      \expandafter\def\csname LT5\endcsname{\color[rgb]{1,1,0}}%
      \expandafter\def\csname LT6\endcsname{\color[rgb]{0,0,0}}%
      \expandafter\def\csname LT7\endcsname{\color[rgb]{1,0.3,0}}%
      \expandafter\def\csname LT8\endcsname{\color[rgb]{0.5,0.5,0.5}}%
    \else
      % gray
      \def\colorrgb#1{\color{black}}%
      \def\colorgray#1{\color[gray]{#1}}%
      \expandafter\def\csname LTw\endcsname{\color{white}}%
      \expandafter\def\csname LTb\endcsname{\color{black}}%
      \expandafter\def\csname LTa\endcsname{\color{black}}%
      \expandafter\def\csname LT0\endcsname{\color{black}}%
      \expandafter\def\csname LT1\endcsname{\color{black}}%
      \expandafter\def\csname LT2\endcsname{\color{black}}%
      \expandafter\def\csname LT3\endcsname{\color{black}}%
      \expandafter\def\csname LT4\endcsname{\color{black}}%
      \expandafter\def\csname LT5\endcsname{\color{black}}%
      \expandafter\def\csname LT6\endcsname{\color{black}}%
      \expandafter\def\csname LT7\endcsname{\color{black}}%
      \expandafter\def\csname LT8\endcsname{\color{black}}%
    \fi
  \fi
  \setlength{\unitlength}{0.0500bp}%
  \begin{picture}(4860.00,3400.00)%
    \gplgaddtomacro\gplbacktext{%
      \csname LTb\endcsname%
      \put(369,1793){\makebox(0,0)[r]{\strut{}-10}}%
      \csname LTb\endcsname%
      \put(369,2317){\makebox(0,0)[r]{\strut{} 0}}%
      \csname LTb\endcsname%
      \put(369,2841){\makebox(0,0)[r]{\strut{} 10}}%
      \csname LTb\endcsname%
      \put(587,1607){\makebox(0,0){\strut{} }}%
      \csname LTb\endcsname%
      \put(819,1607){\makebox(0,0){\strut{} }}%
      \csname LTb\endcsname%
      \put(1052,1607){\makebox(0,0){\strut{} }}%
      \csname LTb\endcsname%
      \put(1284,1607){\makebox(0,0){\strut{} }}%
      \csname LTb\endcsname%
      \put(1516,1607){\makebox(0,0){\strut{} }}%
      \csname LTb\endcsname%
      \put(1748,1607){\makebox(0,0){\strut{} }}%
      \csname LTb\endcsname%
      \put(1981,1607){\makebox(0,0){\strut{} }}%
      \csname LTb\endcsname%
      \put(2213,1607){\makebox(0,0){\strut{} }}%
      \csname LTb\endcsname%
      \put(72,2317){\rotatebox{-270}{\makebox(0,0){\strut{}$b_{1}^j$}}}%
    }%
    \gplgaddtomacro\gplfronttext{%
      \csname LTb\endcsname%
      \put(657,2684){\makebox(0,0){\strut{}(a)}}%
    }%
    \gplgaddtomacro\gplbacktext{%
      \csname LTb\endcsname%
      \put(369,558){\makebox(0,0)[r]{\strut{}-5}}%
      \csname LTb\endcsname%
      \put(369,1083){\makebox(0,0)[r]{\strut{} 0}}%
      \csname LTb\endcsname%
      \put(369,1607){\makebox(0,0)[r]{\strut{} 5}}%
      \csname LTb\endcsname%
      \put(471,372){\makebox(0,0){\strut{}-50}}%
      \csname LTb\endcsname%
      \put(1052,372){\makebox(0,0){\strut{}0}}%
      \csname LTb\endcsname%
      \put(1632,372){\makebox(0,0){\strut{}50}}%
      \csname LTb\endcsname%
      \put(2213,372){\makebox(0,0){\strut{}100}}%
      \csname LTb\endcsname%
      \put(72,1082){\rotatebox{-270}{\makebox(0,0){\strut{}$b_{2}^j$}}}%
      \csname LTb\endcsname%
      \put(1400,93){\makebox(0,0){\strut{}t}}%
    }%
    \gplgaddtomacro\gplfronttext{%
      \csname LTb\endcsname%
      \put(657,1450){\makebox(0,0){\strut{}(b)}}%
    }%
    \gplgaddtomacro\gplbacktext{%
      \csname LTb\endcsname%
      \put(2847,1793){\makebox(0,0)[r]{\strut{}-5}}%
      \csname LTb\endcsname%
      \put(2847,2317){\makebox(0,0)[r]{\strut{} 0}}%
      \csname LTb\endcsname%
      \put(2847,2841){\makebox(0,0)[r]{\strut{} 5}}%
      \csname LTb\endcsname%
      \put(2949,1607){\makebox(0,0){\strut{} }}%
      \csname LTb\endcsname%
      \put(3530,1607){\makebox(0,0){\strut{} }}%
      \csname LTb\endcsname%
      \put(4111,1607){\makebox(0,0){\strut{} }}%
      \csname LTb\endcsname%
      \put(4692,1607){\makebox(0,0){\strut{} }}%
      \csname LTb\endcsname%
      \put(2550,2317){\rotatebox{-270}{\makebox(0,0){\strut{}$b_{3}^j$}}}%
    }%
    \gplgaddtomacro\gplfronttext{%
      \csname LTb\endcsname%
      \put(3135,2684){\makebox(0,0){\strut{}(c)}}%
    }%
    \gplgaddtomacro\gplbacktext{%
      \csname LTb\endcsname%
      \put(2847,558){\makebox(0,0)[r]{\strut{}-5}}%
      \csname LTb\endcsname%
      \put(2847,1064){\makebox(0,0)[r]{\strut{} 0}}%
      \csname LTb\endcsname%
      \put(2847,1570){\makebox(0,0)[r]{\strut{} 5}}%
      \csname LTb\endcsname%
      \put(2949,372){\makebox(0,0){\strut{}-50}}%
      \csname LTb\endcsname%
      \put(3530,372){\makebox(0,0){\strut{}0}}%
      \csname LTb\endcsname%
      \put(4111,372){\makebox(0,0){\strut{}50}}%
      \csname LTb\endcsname%
      \put(4692,372){\makebox(0,0){\strut{}100}}%
      \csname LTb\endcsname%
      \put(2550,1064){\rotatebox{-270}{\makebox(0,0){\strut{}$b_{4}^j$}}}%
      \csname LTb\endcsname%
      \put(3878,93){\makebox(0,0){\strut{}t}}%
    }%
    \gplgaddtomacro\gplfronttext{%
      \csname LTb\endcsname%
      \put(3135,1418){\makebox(0,0){\strut{}(d)}}%
    }%
    \gplbacktext
    \put(0,0){\includegraphics{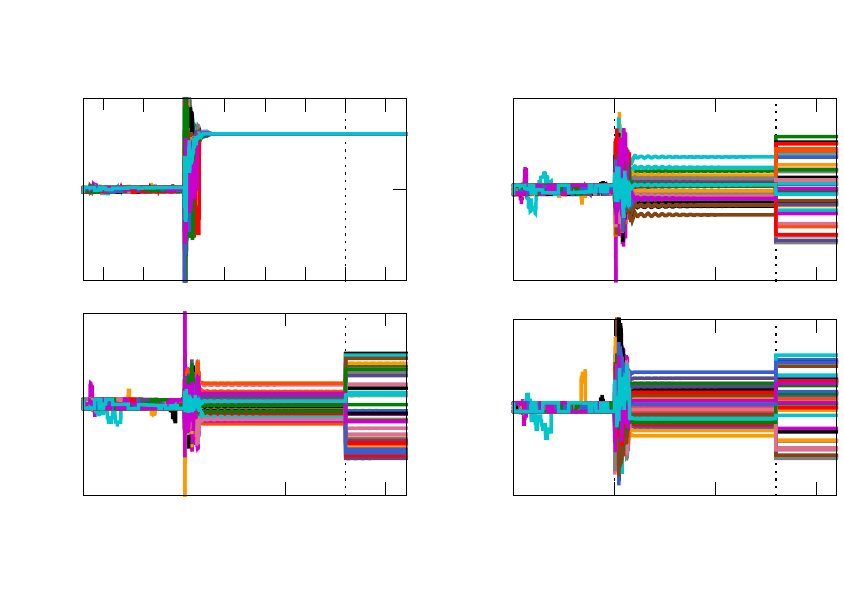}}%
    \gplfronttext
  \end{picture}%
\endgroup
%\input{zeitserieN40m8}
%\input{./b_zeitserie_neu.tex}
%~topology/koeffizienten $ plot_koeffb_paper.sh b_zeitserie_neu 8  40
%~/topology/koeffizienten $ 
\vspace{-0.2cm}
  \caption{ \raggedright
 Control of an 8-cluster state: Fourier coefficients as defined in Eq.~\eqref{eq:coeff}: (a) $b_1^j$, (b) $b_2^j$,
 (c) $b_3^j$, (d)  $b_4^j$. At $t=80$, $b_2^j$, $b_3^j$, $b_4^j$ are
 set to random values.  Parameters as in Fig.~\ref{fig:network8}.}
  \label{fig:b_zeitserie}
\end{figure}

Insight into the network structure can be obtained by a row-wise
discrete Fourier transform of the coupling matrix after successful
control. To do so, we introduce  the following $N \times M$ matrix
\begin{eqnarray} \Gamma_{jk}=\sum_{l=0}^{m-1} \tilde G_{j,k+lM},
\end{eqnarray} where $m$ is the number of nodes in one cluster,
i.e, $m=N/M$, and the symbol \textasciitilde ~denotes the final topology
of the network, i.e., $\tilde G=G(t_{\infty})$. Here, we label the nodes such that the
synchronized state can be described by: 
\begin{eqnarray} \label{eq:sync_state}
r_j&\equiv &r_0 ,\nonumber \\
\varphi_j&\equiv&\Omega t+j\frac{2\pi}{M},
\end{eqnarray}
for $j=1,\ldots,N$, where $r_0$ and $\Omega$ denote the common radius
and the common frequency, respectively. 
 Thus, $\Gamma_{jn}$ represents the total input which node $j$
receives from all nodes in cluster $n$.
We now represent each row of $\Gamma$ as a discrete Fourier series
\begin{eqnarray} \label{eq:ft}
\Gamma_{jk}&=&\frac{b_0^j}{2\cos(\Omega \tau)}+\sum_{l=1}^{N_0-1}
a_l^j\sin(\Phi_{l,k,j})+
b_l^j\cos(\Phi_{l,k,j}) \nonumber \\ 
&& \underbrace{+ \frac{ b_{N_0}^j}{
    \cos(\Omega \tau)^2}\cos(\Phi_{N_0,k,j})}_{\mbox{if } M \mbox{ is even}}
\end{eqnarray}
with the abbreviation $\Phi_{l,k,j}=l(k-j)2\pi/M-\Omega \tau$.  $N_0=(M+1)/2$ if $M$ is odd and $N_0=M/2$ if $M$ is even. 
The coefficients are given by 
\begin{align} \label{eq:coeff}
a_l^j&=&\frac{2}{M}\sum_{k=0}^{M-1}\Gamma_{jk}\sin(\Phi_{l,k,j})=\frac{2}{N}\sum_{k=0}^{N-1}\tilde
G_{jk}\sin(\Phi_{l,k,j})\nonumber \\
b_l^j&=&\frac{2}{M}\sum_{k=0}^{M-1}\Gamma_{jk}\cos(\Phi_{l,k,j})=\frac{2}{N}\sum_{k=0}^{N-1}\tilde
G_{jk}\cos(\Phi_{l,k,j}).
\end{align}
From the constant row sum condition, i.e.,
$\sum_{k=0}^{M-1}\Gamma_{jk}=\sum_{k=0}^{N-1}G_{jk}=1$,
$b_0^j=\frac{1}{2M\cos(\Omega \tau)}$ follows.  $\bar G(n\frac{2
  \pi}{M})$ can be expressed as  $\bar G(n\frac{2 \pi}{M})=\langle \sum_{j=0}^{N-1} 
\Gamma_{j,(j+n)\bmod M} \rangle$.

Next, we discuss the necessary condition for $a_l^j$ and $b_l^j$ such 
that an $M$-cluster state exists as a solution of  Eq.~\eqref{eq:sl_r}.
Substituting Eq.~\eqref{eq:sync_state} into Eqs.~\eqref{eq:sl_r} yields the following
conditions for $r_0$ and $\Omega$:
\begin{subequations} \label{eq:cond0}
\begin{eqnarray} 
r_0^2&=&\lambda+K \sum_{k=0}^{N-1} \tilde  G_{jk} \left[\cos(\Phi_{1,k,j})-1\right],\\
\Omega&=&\omega+K\sum_{k=0}^{M-1}  \tilde G_{jk}  \sin(\Phi_{1,k,j}).
\end{eqnarray}
\end{subequations}
Using Eq.~\eqref{eq:coeff} this can be rewritten as:
\begin{subequations}\label{eq:cond}
\begin{eqnarray} 
r_0^2&=&\lambda+K\left[\frac{b_1^jN}{2}-1\right], \\
\Omega&=&\omega+K\left[\frac{a_1^jN}{2}\right].
\end{eqnarray}
\end{subequations}

Equation \eqref{eq:cond} has  to be fulfilled for all
$j=1,\ldots,N$. Thus, only if $a_1^1=a_1^2=\ldots=a_1^N \equiv a$ and
$b_1^1=b_1^2=\ldots=b_1^N \equiv b$ a solution with a common radius and a
common frequency exists. Note that  there is no restriction on the higher
Fourier coefficients, i.e., on  $a_l^j$ and  $b_l^j$ with $l>1$.

\subsection{ Stability of cluster states } \label{sec:stab-clust-stat}
So far we have discussed the existence of the solution, but not its
stability. Unfortunately, it is not possible to carry out a
systematic stability analysis in all Fourier coefficients as their
number is too high for large $M$. However, there is evidence that the higher Fourier
coefficients do not affect the stability. Figure~\ref{fig:b_zeitserie} shows
the $b_l^j$,$j=1,\ldots, N$, $l=1,\ldots, N_0$ corresponding to the
simulation shown in Fig.~\ref{fig:zeitserie1}. Clearly, the $b_1^j$
converge to a common value $b$ after the control has been switched on at $t=0$ assuring the
existence of a common radius according to Eq.~\eqref{eq:cond}. In
contrast, the higher Fourier coefficients do not approach each other.
At $t=80$, we set each of the higher Fourier coefficients to a random
value in the interval $[-3,3]$ \footnote{We chose this interval for convenience of depiction. Qualitatively, the result is the same for much
larger random numbers.}. As expected  and apparent in
Figs.~\ref{fig:zeitserie1} and~\ref{fig:b_zeitserie}, $a$ and $b$, and thus the
common frequency and radius, do not change
as a result of this perturbation. The higher Fourier coefficients stay at
their new values, since there is no need for them to readjust.
In conclusion, the existence and stability of the cluster solution
does not depend on the higher Fourier coefficients. Thus, the final
values of these higher coefficients
follow from the random initial conditions and are therefore random
themselves. As a consequence they average out, while, on average, the
first Fourier coefficients dominate the topology. In fact, the average
topology is mainly given by $b$ as $a$  and $b_0=\frac{1}{N}$ are
typically small. Thus,  $\bar G(n\frac{2 \pi}{M})=\langle \sum_{j=0}^{N-1} 
\Gamma_{j,(j+n)\bmod M} \rangle \approx b \cos(\frac{2\pi n}{M}-\Omega \tau)$ explaining the
cosine form of the curves shown in Fig.~\ref{weightvsd}.

\section{ Choosing a frequency } \label{sec:choosing-frequency-}
\begin{figure}[]  
% GNUPLOT: LaTeX picture with Postscript
\begingroup
  \makeatletter
  \providecommand\color[2][]{%
    \GenericError{(gnuplot) \space\space\space\@spaces}{%
      Package color not loaded in conjunction with
      terminal option `colourtext'%
    }{See the gnuplot documentation for explanation.%
    }{Either use 'blacktext' in gnuplot or load the package
      color.sty in LaTeX.}%
    \renewcommand\color[2][]{}%
  }%
  \providecommand\includegraphics[2][]{%
    \GenericError{(gnuplot) \space\space\space\@spaces}{%
      Package graphicx or graphics not loaded%
    }{See the gnuplot documentation for explanation.%
    }{The gnuplot epslatex terminal needs graphicx.sty or graphics.sty.}%
    \renewcommand\includegraphics[2][]{}%
  }%
  \providecommand\rotatebox[2]{#2}%
  \@ifundefined{ifGPcolor}{%
    \newif\ifGPcolor
    \GPcolortrue
  }{}%
  \@ifundefined{ifGPblacktext}{%
    \newif\ifGPblacktext
    \GPblacktexttrue
  }{}%
  % define a \g@addto@macro without @ in the name:
  \let\gplgaddtomacro\g@addto@macro
  % define empty templates for all commands taking text:
  \gdef\gplbacktext{}%
  \gdef\gplfronttext{}%
  \makeatother
  \ifGPblacktext
    % no textcolor at all
    \def\colorrgb#1{}%
    \def\colorgray#1{}%
  \else
    % gray or color?
    \ifGPcolor
      \def\colorrgb#1{\color[rgb]{#1}}%
      \def\colorgray#1{\color[gray]{#1}}%
      \expandafter\def\csname LTw\endcsname{\color{white}}%
      \expandafter\def\csname LTb\endcsname{\color{black}}%
      \expandafter\def\csname LTa\endcsname{\color{black}}%
      \expandafter\def\csname LT0\endcsname{\color[rgb]{1,0,0}}%
      \expandafter\def\csname LT1\endcsname{\color[rgb]{0,1,0}}%
      \expandafter\def\csname LT2\endcsname{\color[rgb]{0,0,1}}%
      \expandafter\def\csname LT3\endcsname{\color[rgb]{1,0,1}}%
      \expandafter\def\csname LT4\endcsname{\color[rgb]{0,1,1}}%
      \expandafter\def\csname LT5\endcsname{\color[rgb]{1,1,0}}%
      \expandafter\def\csname LT6\endcsname{\color[rgb]{0,0,0}}%
      \expandafter\def\csname LT7\endcsname{\color[rgb]{1,0.3,0}}%
      \expandafter\def\csname LT8\endcsname{\color[rgb]{0.5,0.5,0.5}}%
    \else
      % gray
      \def\colorrgb#1{\color{black}}%
      \def\colorgray#1{\color[gray]{#1}}%
      \expandafter\def\csname LTw\endcsname{\color{white}}%
      \expandafter\def\csname LTb\endcsname{\color{black}}%
      \expandafter\def\csname LTa\endcsname{\color{black}}%
      \expandafter\def\csname LT0\endcsname{\color{black}}%
      \expandafter\def\csname LT1\endcsname{\color{black}}%
      \expandafter\def\csname LT2\endcsname{\color{black}}%
      \expandafter\def\csname LT3\endcsname{\color{black}}%
      \expandafter\def\csname LT4\endcsname{\color{black}}%
      \expandafter\def\csname LT5\endcsname{\color{black}}%
      \expandafter\def\csname LT6\endcsname{\color{black}}%
      \expandafter\def\csname LT7\endcsname{\color{black}}%
      \expandafter\def\csname LT8\endcsname{\color{black}}%
    \fi
  \fi
  \setlength{\unitlength}{0.0500bp}%
  \begin{picture}(4860.00,3400.00)%
    \gplgaddtomacro\gplbacktext{%
      \csname LTb\endcsname%
      \put(369,1793){\makebox(0,0)[r]{\strut{} 0}}%
      \csname LTb\endcsname%
      \put(369,2142){\makebox(0,0)[r]{\strut{} 1}}%
      \csname LTb\endcsname%
      \put(369,2492){\makebox(0,0)[r]{\strut{} 2}}%
      \csname LTb\endcsname%
      \put(369,2841){\makebox(0,0)[r]{\strut{} 3}}%
      \csname LTb\endcsname%
      \put(471,1607){\makebox(0,0){\strut{} }}%
      \csname LTb\endcsname%
      \put(781,1607){\makebox(0,0){\strut{} }}%
      \csname LTb\endcsname%
      \put(1090,1607){\makebox(0,0){\strut{} }}%
      \csname LTb\endcsname%
      \put(1400,1607){\makebox(0,0){\strut{} }}%
      \csname LTb\endcsname%
      \put(1710,1607){\makebox(0,0){\strut{} }}%
      \csname LTb\endcsname%
      \put(2019,1607){\makebox(0,0){\strut{} }}%
      \csname LTb\endcsname%
      \put(2329,1607){\makebox(0,0){\strut{} }}%
      \csname LTb\endcsname%
      \put(72,2317){\rotatebox{-270}{\makebox(0,0){\strut{}$r_j$}}}%
    }%
    \gplgaddtomacro\gplfronttext{%
      \csname LTb\endcsname%
      \put(2143,2684){\makebox(0,0){\strut{}(a)}}%
    }%
    \gplgaddtomacro\gplbacktext{%
      \csname LTb\endcsname%
      \put(369,558){\makebox(0,0)[r]{\strut{}0}}%
      \csname LTb\endcsname%
      \put(369,1083){\makebox(0,0)[r]{\strut{}$\pi$}}%
      \csname LTb\endcsname%
      \put(369,1607){\makebox(0,0)[r]{\strut{}$2\pi$}}%
      \csname LTb\endcsname%
      \put(781,372){\makebox(0,0){\strut{}0}}%
      \csname LTb\endcsname%
      \put(1400,372){\makebox(0,0){\strut{}40}}%
      \csname LTb\endcsname%
      \put(2019,372){\makebox(0,0){\strut{}80}}%
      \csname LTb\endcsname%
      \put(72,1082){\rotatebox{-270}{\makebox(0,0){\strut{}$\varphi_j$}}}%
      \csname LTb\endcsname%
      \put(1400,93){\makebox(0,0){\strut{}t}}%
    }%
    \gplgaddtomacro\gplfronttext{%
      \csname LTb\endcsname%
      \put(2143,1450){\makebox(0,0){\strut{}(b)}}%
    }%
    \gplgaddtomacro\gplbacktext{%
      \csname LTb\endcsname%
      \put(2847,1793){\makebox(0,0)[r]{\strut{}-20}}%
      \csname LTb\endcsname%
      \put(2847,2055){\makebox(0,0)[r]{\strut{}-10}}%
      \csname LTb\endcsname%
      \put(2847,2317){\makebox(0,0)[r]{\strut{}0}}%
      \csname LTb\endcsname%
      \put(2847,2579){\makebox(0,0)[r]{\strut{}10}}%
      \csname LTb\endcsname%
      \put(2847,2841){\makebox(0,0)[r]{\strut{}20}}%
      \csname LTb\endcsname%
      \put(3259,1607){\makebox(0,0){\strut{} }}%
      \csname LTb\endcsname%
      \put(3879,1607){\makebox(0,0){\strut{} }}%
      \csname LTb\endcsname%
      \put(4498,1607){\makebox(0,0){\strut{} }}%
      \csname LTb\endcsname%
      \put(2448,2317){\rotatebox{-270}{\makebox(0,0){\strut{}$G_{jn}$}}}%
    }%
    \gplgaddtomacro\gplfronttext{%
      \csname LTb\endcsname%
      \put(4622,2684){\makebox(0,0){\strut{}(c)}}%
    }%
    \gplgaddtomacro\gplbacktext{%
      \csname LTb\endcsname%
      \put(2847,558){\makebox(0,0)[r]{\strut{}0}}%
      \csname LTb\endcsname%
      \put(2847,811){\makebox(0,0)[r]{\strut{}1}}%
      \csname LTb\endcsname%
      \put(2847,1064){\makebox(0,0)[r]{\strut{}2}}%
      \csname LTb\endcsname%
      \put(2847,1317){\makebox(0,0)[r]{\strut{}3}}%
      \csname LTb\endcsname%
      \put(2847,1570){\makebox(0,0)[r]{\strut{}4}}%
      \csname LTb\endcsname%
      \put(3259,372){\makebox(0,0){\strut{}0}}%
      \csname LTb\endcsname%
      \put(3879,372){\makebox(0,0){\strut{}40}}%
      \csname LTb\endcsname%
      \put(4498,372){\makebox(0,0){\strut{}80}}%
      \csname LTb\endcsname%
      \put(2652,1064){\rotatebox{-270}{\makebox(0,0){\strut{}$Q_{12},q_{12}, \Omega $}}}%
      \csname LTb\endcsname%
      \put(3878,93){\makebox(0,0){\strut{}t}}%
    }%
    \gplgaddtomacro\gplfronttext{%
      \csname LTb\endcsname%
      \put(4335,1072){\makebox(0,0)[r]{\strut{}$Q_{12}$}}%
      \csname LTb\endcsname%
      \put(4335,886){\makebox(0,0)[r]{\strut{}$q_{12}$}}%
      \csname LTb\endcsname%
      \put(4335,700){\makebox(0,0)[r]{\strut{}$ \Omega $}}%
      \csname LTb\endcsname%
      \put(4622,1418){\makebox(0,0){\strut{}(d)}}%
    }%
    \gplbacktext
    \put(0,0){\includegraphics{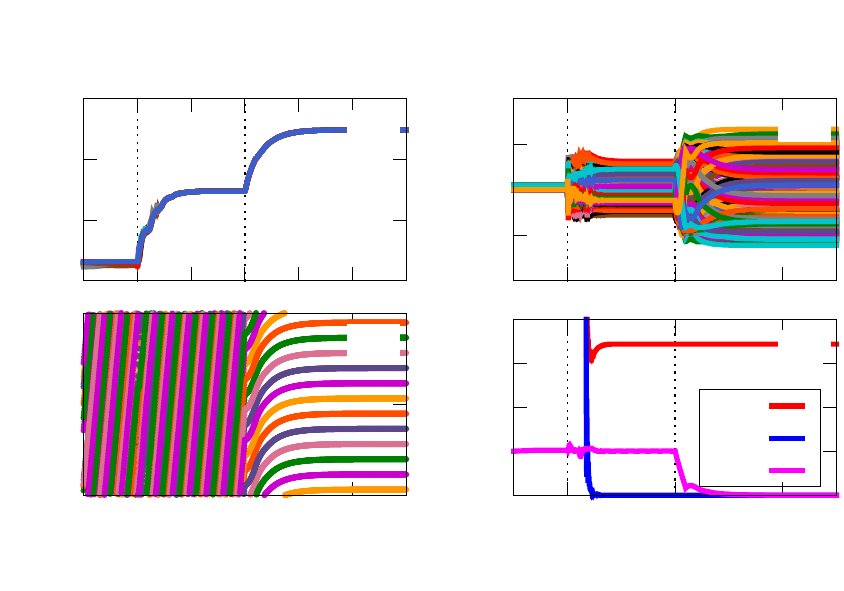}}%
    \gplfronttext
  \end{picture}%
\endgroup
\vspace{-0.2cm}
  \caption{ \raggedright
Oscillation quenching of a splay  state: (a) radii $r_j$, (b) phases $\varphi_j$,
    (c) coupling weights $G_{ij}$, and (d) goal
    function $Q_{12}$, its reduced part $q_{12}$ and the collective
    frequency $\Omega$. $N=12$, $M=12$. At $t=40$ the adaptive control
    is switched off and $a$ is set to $a=\frac{2 \omega}{N K}$. Other parameters as in Fig.~\ref{fig:network8}.}
  \label{fig:gestoppt}
\end{figure}
The representation of the coupling matrix in a Fourier series as in
Eq.~\eqref{eq:ft} is particularly convenient if one wants to choose
a common frequency via constructing an appropriate matrix. To select the
common frequency $\Omega_0$, we set $a=\frac{2}{N}(\frac{\Omega_0-\omega }{K})$  according to
Eq.~\eqref{eq:cond}(b). As an example, we show in
Fig.~\ref{fig:gestoppt} the quenching of oscillations in a splay
state, i.e., we tune $\Omega_0$ to zero: At $t=0$ we start the
adaptive control with $M=N$, i.e., with the goal function leading to a splay state. At $t=40$ the
adaptive control is switched off and $a$ is kept fixed at
$a=\frac{2 \omega}{N K}$ forcing $\Omega$ to approach zero. As a result of
$\Omega_0$ decaying to zero while $a$ is fixed, the coupling weights slightly change after
$t=40$ even though the adaptive control is switched off.

\section{ Control of a subset of links } \label{sec:control-subset-links}
\begin{figure}[]
  \centering
\subfloat{\includegraphics[width=.4\linewidth]{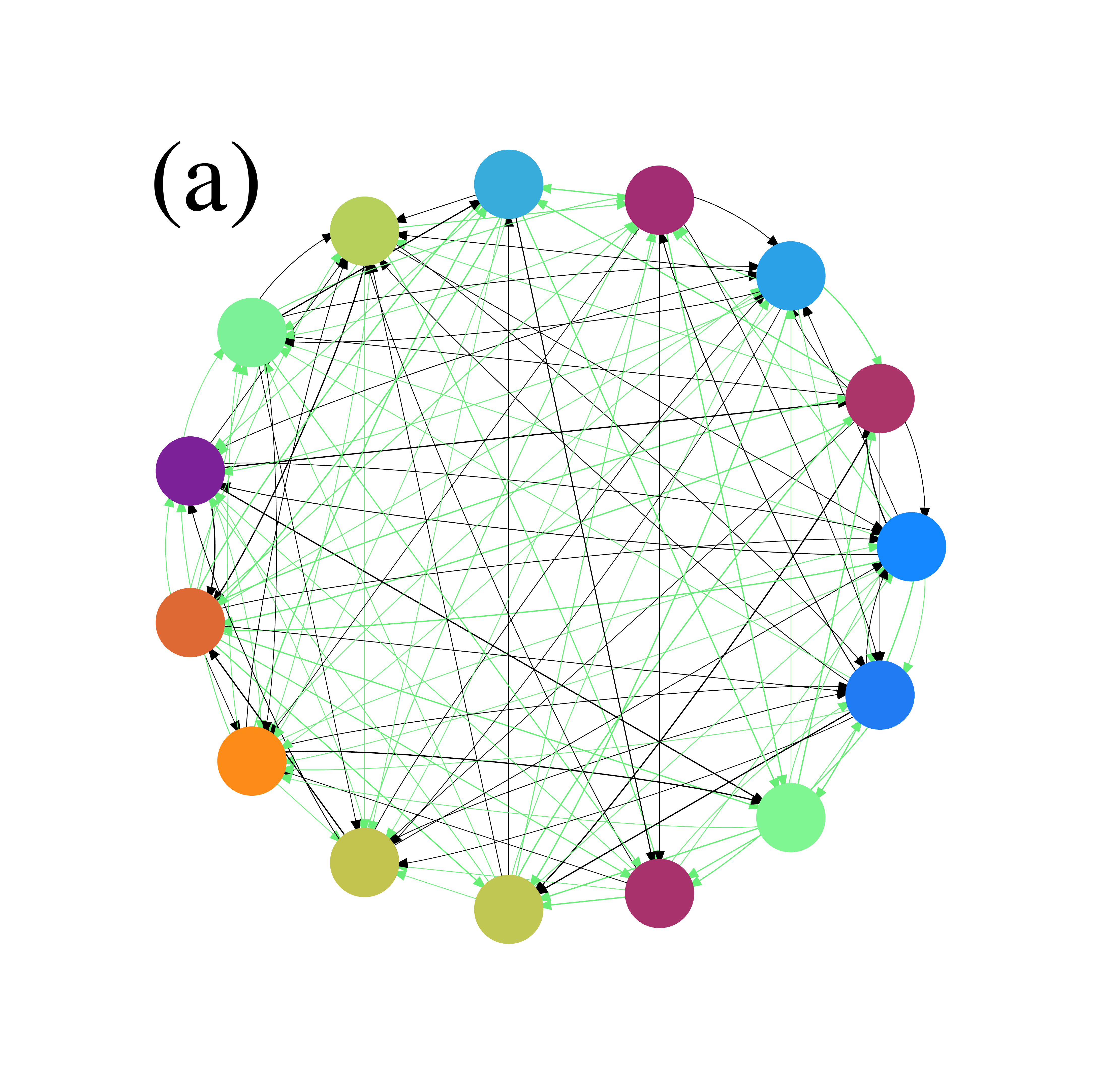}}
\subfloat{\includegraphics[width=.4\linewidth]{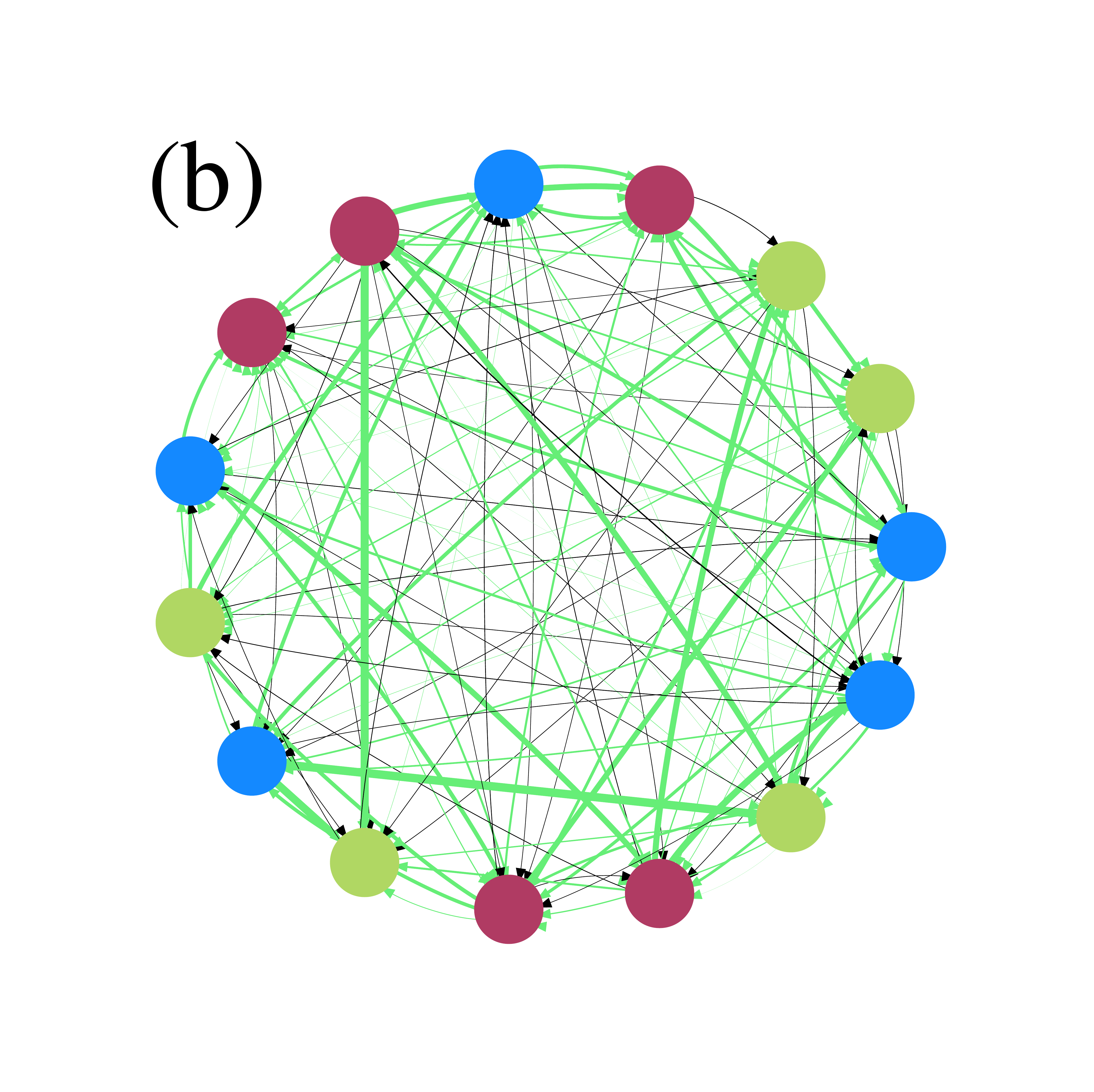}}
 \subfloat{
 \raisebox{-.7cm}{\adjustbox{trim={.3\width} {.15\height} {0.25\width} {.1\height},clip}{
  \includegraphics[width=.3\linewidth,
  height=.55\linewidth]{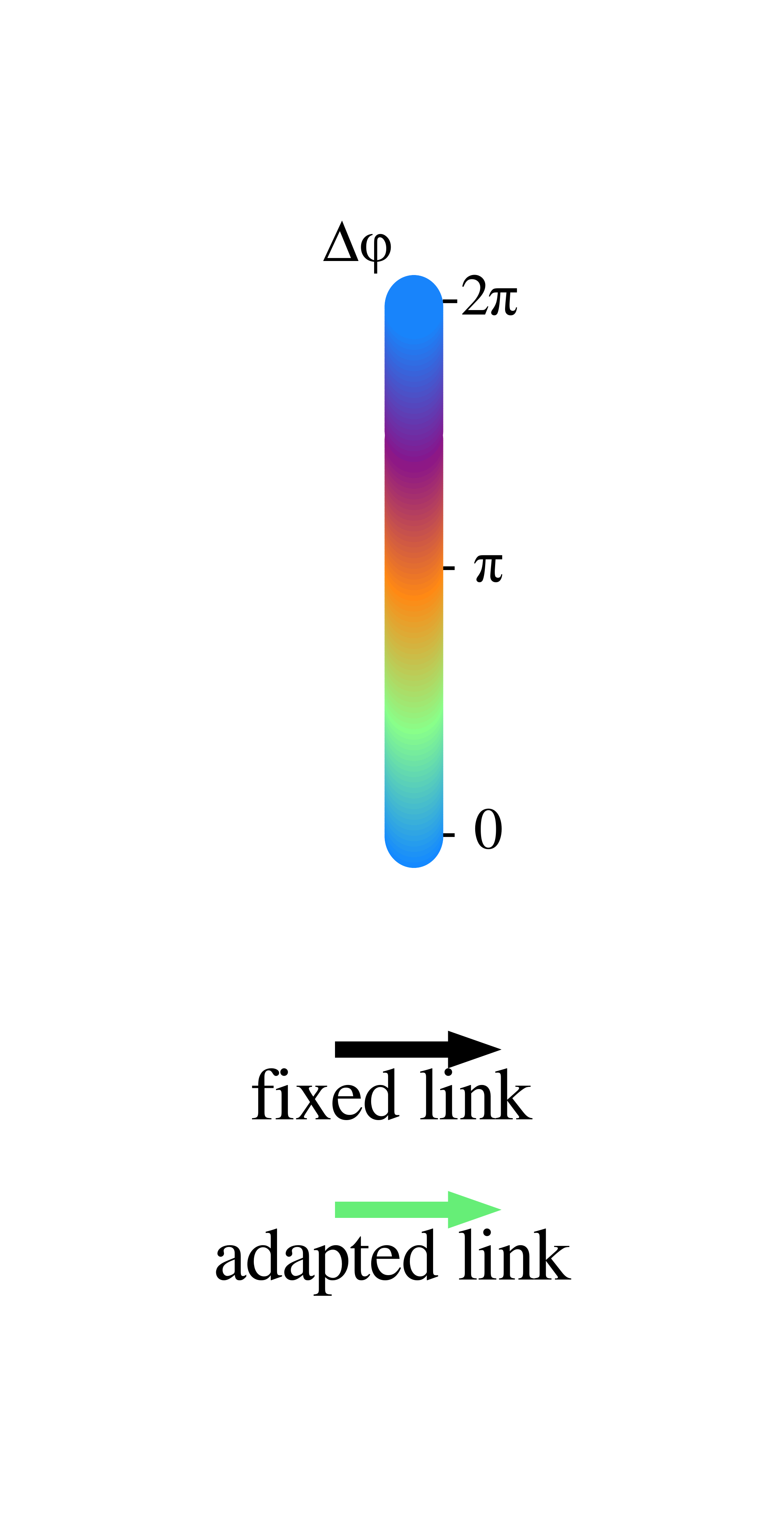}}}}
% trim=l b r t
\vspace{-0.2cm}
  \caption{\raggedright  Topology (a) before ($t=0$)  and (b) after
    control ($t=t_c$). Black links:
    fixed links; green (gray): adapted links. Color code of nodes: Phase difference
    $\Delta \varphi=\varphi_j-\varphi_0$ with respect to
    the first node. Parameters: $K=0.1$, $\tau=3$, $\gamma_G=10$, $A/L=0.3$, $P/L=0.4$, $N=15$,
$M=3$. Other parameters as in Fig.~\ref{fig:network8}. Initial
conditions: directed random network.
}
  \label{fig:netzwerk}
\end{figure}

So far we have controlled every link of the network. However, this is not
necessary. It is sufficient to control a subset of links, while the
other links are left fixed. We demonstrate this with  an example of a
directed random network constructed of $P$ links, which
are chosen from the $L=N(N-1)$ possible links excluding
self-coupling. From these $P$ links, we select, again randomly, $A$
links
which are subject to adaptation as given by
Eq.~\eqref{controlalgo}. As an example, Fig.~\ref{fig:netzwerk}
depicts the generation of a 3-cluster state in a network of 15 notes.
Figure~\ref{fig:netzwerk}(a) shows the network before the control:
Black links mark the $P-A$ fixed links, green (gray) links depict
the $A$ links that will be adapted. Obviously, only these links change
their strength during the adaptation process as can be seen in
Fig.~\ref{fig:netzwerk}(b), which displays the network after it
has reached the desired 3-cluster state and the change of topology
terminates. 

\begin{figure}[]
% GNUPLOT: LaTeX picture with Postscript
\begingroup
  \makeatletter
  \providecommand\color[2][]{%
    \GenericError{(gnuplot) \space\space\space\@spaces}{%
      Package color not loaded in conjunction with
      terminal option `colourtext'%
    }{See the gnuplot documentation for explanation.%
    }{Either use 'blacktext' in gnuplot or load the package
      color.sty in LaTeX.}%
    \renewcommand\color[2][]{}%
  }%
  \providecommand\includegraphics[2][]{%
    \GenericError{(gnuplot) \space\space\space\@spaces}{%
      Package graphicx or graphics not loaded%
    }{See the gnuplot documentation for explanation.%
    }{The gnuplot epslatex terminal needs graphicx.sty or graphics.sty.}%
    \renewcommand\includegraphics[2][]{}%
  }%
  \providecommand\rotatebox[2]{#2}%
  \@ifundefined{ifGPcolor}{%
    \newif\ifGPcolor
    \GPcolortrue
  }{}%
  \@ifundefined{ifGPblacktext}{%
    \newif\ifGPblacktext
    \GPblacktexttrue
  }{}%
  % define a \g@addto@macro without @ in the name:
  \let\gplgaddtomacro\g@addto@macro
  % define empty templates for all commands taking text:
  \gdef\gplbacktext{}%
  \gdef\gplfronttext{}%
  \makeatother
  \ifGPblacktext
    % no textcolor at all
    \def\colorrgb#1{}%
    \def\colorgray#1{}%
  \else
    % gray or color?
    \ifGPcolor
      \def\colorrgb#1{\color[rgb]{#1}}%
      \def\colorgray#1{\color[gray]{#1}}%
      \expandafter\def\csname LTw\endcsname{\color{white}}%
      \expandafter\def\csname LTb\endcsname{\color{black}}%
      \expandafter\def\csname LTa\endcsname{\color{black}}%
      \expandafter\def\csname LT0\endcsname{\color[rgb]{1,0,0}}%
      \expandafter\def\csname LT1\endcsname{\color[rgb]{0,1,0}}%
      \expandafter\def\csname LT2\endcsname{\color[rgb]{0,0,1}}%
      \expandafter\def\csname LT3\endcsname{\color[rgb]{1,0,1}}%
      \expandafter\def\csname LT4\endcsname{\color[rgb]{0,1,1}}%
      \expandafter\def\csname LT5\endcsname{\color[rgb]{1,1,0}}%
      \expandafter\def\csname LT6\endcsname{\color[rgb]{0,0,0}}%
      \expandafter\def\csname LT7\endcsname{\color[rgb]{1,0.3,0}}%
      \expandafter\def\csname LT8\endcsname{\color[rgb]{0.5,0.5,0.5}}%
    \else
      % gray
      \def\colorrgb#1{\color{black}}%
      \def\colorgray#1{\color[gray]{#1}}%
      \expandafter\def\csname LTw\endcsname{\color{white}}%
      \expandafter\def\csname LTb\endcsname{\color{black}}%
      \expandafter\def\csname LTa\endcsname{\color{black}}%
      \expandafter\def\csname LT0\endcsname{\color{black}}%
      \expandafter\def\csname LT1\endcsname{\color{black}}%
      \expandafter\def\csname LT2\endcsname{\color{black}}%
      \expandafter\def\csname LT3\endcsname{\color{black}}%
      \expandafter\def\csname LT4\endcsname{\color{black}}%
      \expandafter\def\csname LT5\endcsname{\color{black}}%
      \expandafter\def\csname LT6\endcsname{\color{black}}%
      \expandafter\def\csname LT7\endcsname{\color{black}}%
      \expandafter\def\csname LT8\endcsname{\color{black}}%
    \fi
  \fi
  \setlength{\unitlength}{0.0500bp}%
  \begin{picture}(4860.00,3400.00)%
    \gplgaddtomacro\gplbacktext{%
      \csname LTb\endcsname%
      \put(369,1793){\makebox(0,0)[r]{\strut{} 0}}%
      \csname LTb\endcsname%
      \put(369,2212){\makebox(0,0)[r]{\strut{} 0.2}}%
      \csname LTb\endcsname%
      \put(369,2631){\makebox(0,0)[r]{\strut{} 0.4}}%
      \csname LTb\endcsname%
      \put(587,1607){\makebox(0,0){\strut{} }}%
      \csname LTb\endcsname%
      \put(819,1607){\makebox(0,0){\strut{} }}%
      \csname LTb\endcsname%
      \put(1052,1607){\makebox(0,0){\strut{} }}%
      \csname LTb\endcsname%
      \put(1284,1607){\makebox(0,0){\strut{} }}%
      \csname LTb\endcsname%
      \put(1516,1607){\makebox(0,0){\strut{} }}%
      \csname LTb\endcsname%
      \put(1748,1607){\makebox(0,0){\strut{} }}%
      \csname LTb\endcsname%
      \put(1981,1607){\makebox(0,0){\strut{} }}%
      \csname LTb\endcsname%
      \put(2213,1607){\makebox(0,0){\strut{} }}%
    }%
    \gplgaddtomacro\gplfronttext{%
      \csname LTb\endcsname%
      \put(2143,2684){\makebox(0,0){\strut{}(a)}}%
    }%
    \gplgaddtomacro\gplbacktext{%
      \csname LTb\endcsname%
      \put(369,558){\makebox(0,0)[r]{\strut{}0}}%
      \csname LTb\endcsname%
      \put(369,1083){\makebox(0,0)[r]{\strut{}$\pi$}}%
      \csname LTb\endcsname%
      \put(369,1607){\makebox(0,0)[r]{\strut{}$2\pi$}}%
      \csname LTb\endcsname%
      \put(471,372){\makebox(0,0){\strut{}-50}}%
      \csname LTb\endcsname%
      \put(1052,372){\makebox(0,0){\strut{}0}}%
      \csname LTb\endcsname%
      \put(1632,372){\makebox(0,0){\strut{}50}}%
      \csname LTb\endcsname%
      \put(2213,372){\makebox(0,0){\strut{}100}}%
      \csname LTb\endcsname%
      \put(72,1082){\rotatebox{-270}{\makebox(0,0){\strut{}$\varphi_j-\varphi_0$}}}%
      \csname LTb\endcsname%
      \put(1400,93){\makebox(0,0){\strut{}t}}%
    }%
    \gplgaddtomacro\gplfronttext{%
      \csname LTb\endcsname%
      \put(2143,1450){\makebox(0,0){\strut{}(b)}}%
    }%
    \gplgaddtomacro\gplbacktext{%
      \csname LTb\endcsname%
      \put(2847,1968){\makebox(0,0)[r]{\strut{}-2}}%
      \csname LTb\endcsname%
      \put(2847,2317){\makebox(0,0)[r]{\strut{}0}}%
      \csname LTb\endcsname%
      \put(2847,2666){\makebox(0,0)[r]{\strut{}2}}%
      \csname LTb\endcsname%
      \put(2949,1607){\makebox(0,0){\strut{} }}%
      \csname LTb\endcsname%
      \put(3530,1607){\makebox(0,0){\strut{} }}%
      \csname LTb\endcsname%
      \put(4111,1607){\makebox(0,0){\strut{} }}%
      \csname LTb\endcsname%
      \put(4692,1607){\makebox(0,0){\strut{} }}%
      \csname LTb\endcsname%
      \put(2550,2317){\rotatebox{-270}{\makebox(0,0){\strut{}$G_{jn}$}}}%
    }%
    \gplgaddtomacro\gplfronttext{%
      \csname LTb\endcsname%
      \put(4622,2684){\makebox(0,0){\strut{}(c)}}%
    }%
    \gplgaddtomacro\gplbacktext{%
      \csname LTb\endcsname%
      \put(2847,558){\makebox(0,0)[r]{\strut{}0}}%
      \csname LTb\endcsname%
      \put(2847,947){\makebox(0,0)[r]{\strut{}5}}%
      \csname LTb\endcsname%
      \put(2847,1336){\makebox(0,0)[r]{\strut{}10}}%
      \csname LTb\endcsname%
      \put(2949,372){\makebox(0,0){\strut{}-50}}%
      \csname LTb\endcsname%
      \put(3530,372){\makebox(0,0){\strut{}0}}%
      \csname LTb\endcsname%
      \put(4111,372){\makebox(0,0){\strut{}50}}%
      \csname LTb\endcsname%
      \put(4692,372){\makebox(0,0){\strut{}100}}%
      \csname LTb\endcsname%
      \put(2550,1064){\rotatebox{-270}{\makebox(0,0){\strut{}$Q_3,q_3$}}}%
      \csname LTb\endcsname%
      \put(3878,93){\makebox(0,0){\strut{}t}}%
    }%
    \gplgaddtomacro\gplfronttext{%
      \csname LTb\endcsname%
      \put(4056,1376){\makebox(0,0)[r]{\strut{}$Q_3$}}%
      \csname LTb\endcsname%
      \put(4056,1190){\makebox(0,0)[r]{\strut{}$q_3$}}%
      \csname LTb\endcsname%
      \put(4622,1418){\makebox(0,0){\strut{}(d)}}%
    }%
    \gplbacktext
    \put(0,0){\includegraphics{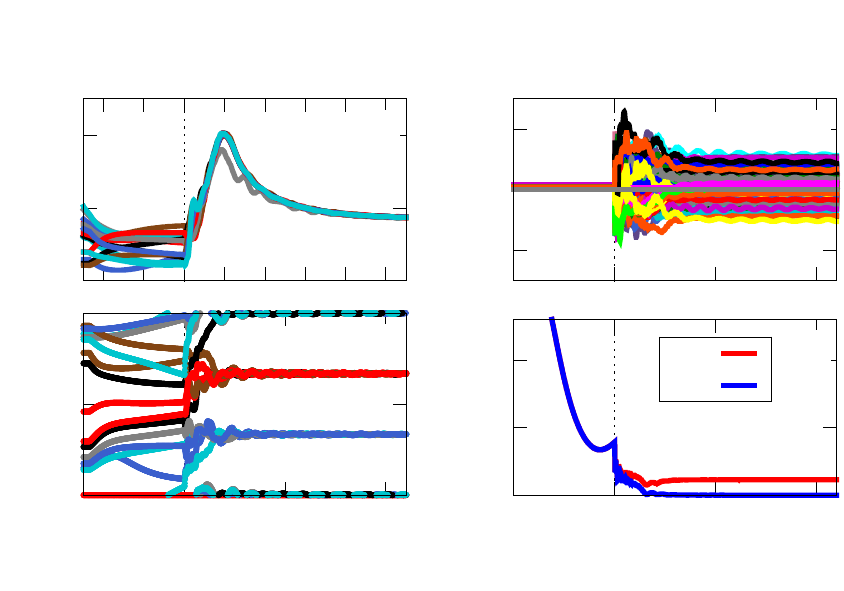}}%
    \gplfronttext
  \end{picture}%
\endgroup
% \input{./zeitserieN15M3}
% trim=l b r t
\vspace{-0.2cm}
  \caption{\raggedright Restricted control of a 3-cluster state: (a)
    radii $r_j$, (b) phase
    difference $\varphi_j-\varphi_0$ with respect to
    the first node, (c) coupling weights $G_{ij}$, and (d) goal
    function $Q_3$ and it its reduced part $q_3$. Parameters as in  Fig.~\ref{fig:netzwerk}.}
  \label{fig:zeitserie}
\end{figure}

The corresponding time series are shown in
Fig.~\ref{fig:zeitserie}(a)-(d) for the radii, the phase
differences, the elements of the coupling matrix, and the goal
function, respectively.  As can be seen in Fig.~\ref{fig:zeitserie}(b)
and (d),
after successful control with goal function $Q_3$  the network consists
of  3 equally sized clusters and the reduced goal function  $q_3$ 
is zero.

\begin{figure}[h]
\centering
% GNUPLOT: LaTeX picture with Postscript
\begingroup
  \makeatletter
  \providecommand\color[2][]{%
    \GenericError{(gnuplot) \space\space\space\@spaces}{%
      Package color not loaded in conjunction with
      terminal option `colourtext'%
    }{See the gnuplot documentation for explanation.%
    }{Either use 'blacktext' in gnuplot or load the package
      color.sty in LaTeX.}%
    \renewcommand\color[2][]{}%
  }%
  \providecommand\includegraphics[2][]{%
    \GenericError{(gnuplot) \space\space\space\@spaces}{%
      Package graphicx or graphics not loaded%
    }{See the gnuplot documentation for explanation.%
    }{The gnuplot epslatex terminal needs graphicx.sty or graphics.sty.}%
    \renewcommand\includegraphics[2][]{}%
  }%
  \providecommand\rotatebox[2]{#2}%
  \@ifundefined{ifGPcolor}{%
    \newif\ifGPcolor
    \GPcolortrue
  }{}%
  \@ifundefined{ifGPblacktext}{%
    \newif\ifGPblacktext
    \GPblacktexttrue
  }{}%
  % define a \g@addto@macro without @ in the name:
  \let\gplgaddtomacro\g@addto@macro
  % define empty templates for all commands taking text:
  \gdef\gplbacktext{}%
  \gdef\gplfronttext{}%
  \makeatother
  \ifGPblacktext
    % no textcolor at all
    \def\colorrgb#1{}%
    \def\colorgray#1{}%
  \else
    % gray or color?
    \ifGPcolor
      \def\colorrgb#1{\color[rgb]{#1}}%
      \def\colorgray#1{\color[gray]{#1}}%
      \expandafter\def\csname LTw\endcsname{\color{white}}%
      \expandafter\def\csname LTb\endcsname{\color{black}}%
      \expandafter\def\csname LTa\endcsname{\color{black}}%
      \expandafter\def\csname LT0\endcsname{\color[rgb]{1,0,0}}%
      \expandafter\def\csname LT1\endcsname{\color[rgb]{0,1,0}}%
      \expandafter\def\csname LT2\endcsname{\color[rgb]{0,0,1}}%
      \expandafter\def\csname LT3\endcsname{\color[rgb]{1,0,1}}%
      \expandafter\def\csname LT4\endcsname{\color[rgb]{0,1,1}}%
      \expandafter\def\csname LT5\endcsname{\color[rgb]{1,1,0}}%
      \expandafter\def\csname LT6\endcsname{\color[rgb]{0,0,0}}%
      \expandafter\def\csname LT7\endcsname{\color[rgb]{1,0.3,0}}%
      \expandafter\def\csname LT8\endcsname{\color[rgb]{0.5,0.5,0.5}}%
    \else
      % gray
      \def\colorrgb#1{\color{black}}%
      \def\colorgray#1{\color[gray]{#1}}%
      \expandafter\def\csname LTw\endcsname{\color{white}}%
      \expandafter\def\csname LTb\endcsname{\color{black}}%
      \expandafter\def\csname LTa\endcsname{\color{black}}%
      \expandafter\def\csname LT0\endcsname{\color{black}}%
      \expandafter\def\csname LT1\endcsname{\color{black}}%
      \expandafter\def\csname LT2\endcsname{\color{black}}%
      \expandafter\def\csname LT3\endcsname{\color{black}}%
      \expandafter\def\csname LT4\endcsname{\color{black}}%
      \expandafter\def\csname LT5\endcsname{\color{black}}%
      \expandafter\def\csname LT6\endcsname{\color{black}}%
      \expandafter\def\csname LT7\endcsname{\color{black}}%
      \expandafter\def\csname LT8\endcsname{\color{black}}%
    \fi
  \fi
  \setlength{\unitlength}{0.0500bp}%
  \begin{picture}(4860.00,3400.00)%
    \gplgaddtomacro\gplbacktext{%
    }%
    \gplgaddtomacro\gplfronttext{%
      \csname LTb\endcsname%
      \put(734,396){\makebox(0,0){\strut{} 0.2}}%
      \csname LTb\endcsname%
      \put(1533,396){\makebox(0,0){\strut{} 0.4}}%
      \csname LTb\endcsname%
      \put(2332,396){\makebox(0,0){\strut{} 0.6}}%
      \csname LTb\endcsname%
      \put(3131,396){\makebox(0,0){\strut{} 0.8}}%
      \csname LTb\endcsname%
      \put(3930,396){\makebox(0,0){\strut{} 1}}%
      \csname LTb\endcsname%
      \put(2332,117){\makebox(0,0){\strut{}$\frac{P}{L}$}}%
      \csname LTb\endcsname%
      \put(602,857){\makebox(0,0)[r]{\strut{} 0.2}}%
      \csname LTb\endcsname%
      \put(602,1295){\makebox(0,0)[r]{\strut{} 0.4}}%
      \csname LTb\endcsname%
      \put(602,1732){\makebox(0,0)[r]{\strut{} 0.6}}%
      \csname LTb\endcsname%
      \put(602,2170){\makebox(0,0)[r]{\strut{} 0.8}}%
      \csname LTb\endcsname%
      \put(602,2608){\makebox(0,0)[r]{\strut{} 1}}%
      \csname LTb\endcsname%
      \put(143,1623){\rotatebox{-270}{\makebox(0,0){\strut{}$\frac{A}{L}$}}}%
      \csname LTb\endcsname%
      \put(4272,638){\makebox(0,0)[l]{\strut{} 0}}%
      \csname LTb\endcsname%
      \put(4272,1032){\makebox(0,0)[l]{\strut{} 0.2}}%
      \csname LTb\endcsname%
      \put(4272,1426){\makebox(0,0)[l]{\strut{} 0.4}}%
      \csname LTb\endcsname%
      \put(4272,1820){\makebox(0,0)[l]{\strut{} 0.6}}%
      \csname LTb\endcsname%
      \put(4272,2214){\makebox(0,0)[l]{\strut{} 0.8}}%
      \csname LTb\endcsname%
      \put(4272,2608){\makebox(0,0)[l]{\strut{} 1}}%
      \csname LTb\endcsname%
      \put(4731,1623){\rotatebox{-270}{\makebox(0,0){\strut{}$f_c$}}}%
    }%
    \gplgaddtomacro\gplbacktext{%
    }%
    \gplgaddtomacro\gplfronttext{%
      \csname LTb\endcsname%
      \put(876,1293){\makebox(0,0){\strut{}}}%
      \csname LTb\endcsname%
      \put(1179,1293){\makebox(0,0){\strut{}}}%
      \csname LTb\endcsname%
      \put(1481,1293){\makebox(0,0){\strut{}}}%
      \csname LTb\endcsname%
      \put(1783,1293){\makebox(0,0){\strut{}}}%
      \csname LTb\endcsname%
      \put(2086,1293){\makebox(0,0){\strut{}}}%
      \csname LTb\endcsname%
      \put(743,1637){\makebox(0,0)[r]{\strut{}}}%
      \csname LTb\endcsname%
      \put(743,1841){\makebox(0,0)[r]{\strut{}}}%
      \csname LTb\endcsname%
      \put(743,2043){\makebox(0,0)[r]{\strut{}}}%
      \csname LTb\endcsname%
      \put(743,2247){\makebox(0,0)[r]{\strut{}}}%
      \csname LTb\endcsname%
      \put(743,2451){\makebox(0,0)[r]{\strut{}}}%
      \csname LTb\endcsname%
      \put(2279,1535){\makebox(0,0)[l]{\strut{} 0}}%
      \csname LTb\endcsname%
      \put(2279,1764){\makebox(0,0)[l]{\strut{} 100}}%
      \csname LTb\endcsname%
      \put(2279,1993){\makebox(0,0)[l]{\strut{} 200}}%
      \csname LTb\endcsname%
      \put(2279,2222){\makebox(0,0)[l]{\strut{} 300}}%
      \csname LTb\endcsname%
      \put(2279,2451){\makebox(0,0)[l]{\strut{} 400}}%
      \csname LTb\endcsname%
      \put(2738,1993){\rotatebox{-270}{\makebox(0,0){\strut{}$t_c$}}}%
    }%
    \gplbacktext
    \put(0,0){\includegraphics{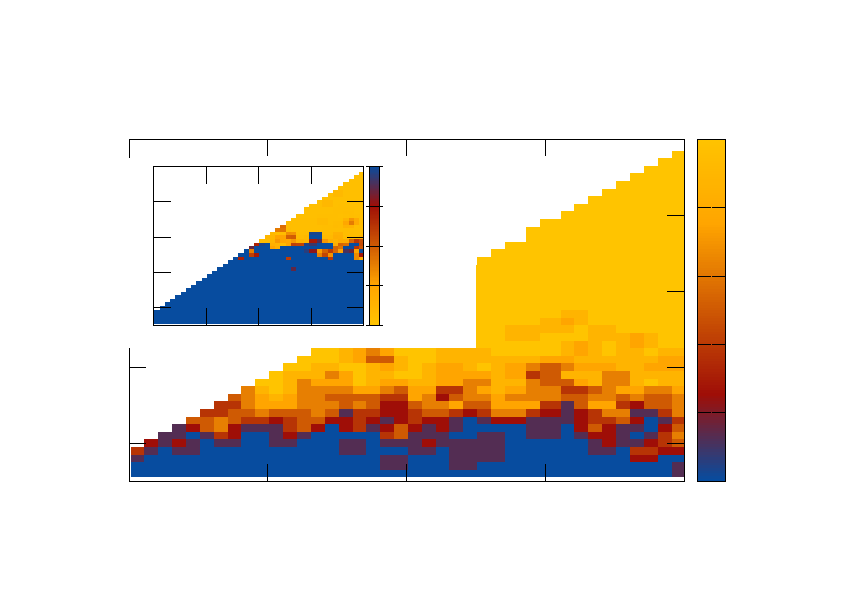}}%
    \gplfronttext
  \end{picture}%
\endgroup
\caption{\raggedright Fraction $f_c$ of successfully controlled
  networks as a function of number of random links $P$ and number of
  controlled links $A$, normalized by $L=N(N-1)$. Inset:
times
$t_c$ needed to reach the control goal. $K=0.2$. We simulated 10 realizations for each parameter combination.
Other parameters as in
Fig.~\ref{fig:netzwerk}.}
\label{fig:fraction}
\end{figure}

We now want to test how successful our method is in dependence upon the
links present in the networks, and upon the fraction of these links subject
to adaptation: Figure~\ref{fig:fraction} shows the fraction $f_c$ of successfully
controlled networks and as an inset, the control time $t_c$ as
function of $P/L$ and $A/L$. We define a network as successfully
controlled in an $M$-cluster state (including the cases of
unequally sized clusters just discussed) at time $t_c$ if it was in this
state for $t \in [t_c-1,t_c]$.

Note that the success rate is fairly independent of the total
number of links $P$ in the network, but depends mainly on the ratio of adapted links to
all possible links. In other words, the links additionally present in
the
network, but not subject to control, have almost no effect on the
synchronizability. If the number of adapted links reaches about $30\%$
the
control still works in more than  $80\%$ of the cases. 

Furthermore, we
study $f_c$ and $t_c$ in dependence on the coupling strength $K$ and
the delay time $\tau$ for a fully connected and adapted network, i.e.,
$P/L=A/L=1$. We find that the fraction of successfully controlled
networks $f_c$ is close to 1 and
$t_c$ is roughly 10 units in the considered range $0.1 <K\le 5$, $0\le\tau\le 3\pi$
(not shown here), demonstrating that our method works very reliably
independently of the coupling parameters.

\section{ Conclusion } \label{sec:conclusion-}
We have proposed a novel method to control cluster synchronization in a
network of time-delay coupled oscillators. Our method uses the speed-gradient algorithm to adapt the topology. The considered goal function
is based on a generalized Kuramoto order
parameter and is independent of the ordering of the nodes. An
additional term ensures amplitude synchronization. The control
scheme is very robust. Numerically, we studied the robustness towards
different initial conditions and its dependence on the overall
coupling strength and delay time.% So far only few control methods were
%developed to control cluster synchronization.

The topology of the network after successful control is
modulated by the coupling  delay. A row-wise discrete Fourier transform of
the coupling matrix gives insight into these delay modulations. Necessary
conditions for the existence of a common radius and a common frequency
give rise to restrictions affecting the first Fourier coefficients,
while there is no restriction for the higher Fourier coefficients.  
We also found that the stability of the cluster states is not affected by
the higher Fourier coefficients. Thus, we conclude that the higher
Fourier coefficients are mainly dependent on the random initial
conditions and are therefore randomly distributed. On
average, the network topology is therefore dominated by the first
Fourier coefficients leading to the observed delay modulation.

We show that we can make use of the first Fourier coefficients in
order to find topologies that lead to cluster states with a given
common frequency. As an example, we quench the oscillations in a
Stuart-Landau oscillator. Thus, we have found a very versatile  method to
construct networks which show a desired dynamical behavior. 

Since in many real-world networks not all links are accessible to
control, we applied the adaptation algorithm to random networks where
we chose a random subset of links which we controlled, while the other
links remained fixed. We found that the control is successful if the
number of adapted links is equal or higher than approximately 30\% of
all possible links, independently of
the number of actual fixed links.  For practical applications this
opens up the
possibility to apply the method more efficiently.
  
Given the paradigmatic nature of the Stuart-Landau oscillator as a
generic model, we expect broad applicability, for instance, to control
synchronization of networks in medicine, chemistry or mechanical
engineering or as self-organizing mechanisms in biological networks.

\begin{acknowledgments}
This work was supported by Deutsche Forschungsgemeinschaft in the
framework of SFB 910. JL acknowledges the
support by the German-Russian Interdisciplinary Science Center (G-RISC) funded by the German Federal
Foreign Office via the German Academic Exchange Service
(DAAD). PH acknowledges support by the Federal Ministry of Education
and Research (BMBF), Germany (grant no. 01GQ1001B). We would like to thank Thomas Isele for fruitful discussion.
\end{acknowledgments}
%\bibliography{ref}
%\bibliographystyle{prwithtitle}
%\bibliographystyle{prsty-fullauthor}

\end{document}